%% file: main.tex
\documentclass[11pt]{article}

\usepackage[letterpaper, left=1in, right=1in, top=1in,bottom=1in]{geometry}

\usepackage{parskip}

\usepackage{booktabs} % For formal tables

\usepackage[utf8]{inputenc}

\usepackage{geometry}
 \newcommand{\Xomit}[1]{}

\usepackage{graphpap,amscd,mathrsfs,graphicx,lscape,enumitem,dsfont,bm,url,subfigure}
\usepackage{epsfig,amstext,xspace}
\usepackage{algorithm,algorithmic,comment}

\usepackage{thmtools,thm-restate}

\newcommand{\E}{\mathbf{E}}

\usepackage{algorithm,algorithmic}
\usepackage{auxiliary}
\usepackage[utf8]{inputenc} % allow utf-8 input
\usepackage[T1]{fontenc}    % use 8-bit T1 fonts
\usepackage{hyperref}       % hyperlinks
\usepackage{url}            % simple URL typesetting
\usepackage{booktabs}       % professional-quality tables
\usepackage{amsfonts}       % blackboard math symbols
\usepackage{nicefrac}       % compact symbols for 1/2, etc.
\usepackage{microtype}      % microtypography

\usepackage{color}              % Need the color package
\usepackage[suppress]{color-edits}
\addauthor{et}{blue}
\addauthor{vs}{red}
\addauthor{tl}{cyan}
\addauthor{tlfp}{green}

\newcommand{\polylog}{\ensuremath{\textnormal{polylog}}}

\newcommand{\M}{\ensuremath{\mathcal{M}}}
\newcommand{\rev}{\ensuremath{\mathcal{R}}}
\renewcommand{\vec}[1]{\ensuremath{{\bf #1}}}

\newcommand{\A}{\ensuremath{\mathcal{A}}}

\newcommand{\V}{\ensuremath{\mathcal{V}}}
\newcommand{\X}{\ensuremath{\mathcal{X}}}

\newcommand{\rsolv}{\ensuremath{{\bf \sigma}}}

\newcommand{\Omitproof}[1]{}

\newcommand{\Omit}[1]{}
\begin{document}
%%%%%%%%%%%%%%%%

\title{Learning and Efficiency in Games with Dynamic Population}

 \author{
 Thodoris Lykouris\thanks{Microsoft Research NYC, \texttt{thlykour@microsoft.com}. Work mostly conducted while the author was a Ph.D. student at Cornell University and was supported by a Google Ph.D. fellowship.}
 \and
Vasilis Syrgkanis\thanks{Microsoft Research NE, \texttt{vasy@microsoft.edu}}
 \and
 \'{E}va Tardos\thanks{Cornell University, \texttt{eva.tardos@cornell.edu}. Work supported in part by NSF grant CCF-1563714, CCF-1408673, ONR grant N00014-08-1-0031, and ASOFR grant F5684A1.}
}
\date{First version: May 2015\\Current version: May 2020%
%\etcomment{see changed footnote}\tlcomment{I was thinking of this text for the difference to the previous ArXiv version not for the letter to them so for ArXiv I would remove the full proofs added as they existed in the previous arxiv but for the letter we should for sure have it}
\footnote{Preliminary version appeared in ACM Symposium on Discrete Algorithms 2016 (SODA 2016). This version adds a major new result: asymptotic optimality of simultaneous second-price auctions with dynamic population. Presentation is significantly simplified %, full proofs added, 
and all results are presented parametrically.}}
\maketitle
\thispagestyle{empty}

\begin{abstract}
\input{abstract}
\end{abstract}

% \KEYWORDS{No-regret learning, price of anarchy, auctions, congestion games, differential privacy} 
% \HISTORY{First version: May 2015. Current version: June 2016. Preliminary version appeared at the 2016 SIAM Symposium on Discrete Algorithms, SODA'16.}

\newpage
\setcounter{page}{1}

\section{Introduction}
\label{sec:intro}
\input{introduction}

\section{Preliminaries}
\label{sec:preliminaries}
\input{preliminaries}

\section{Price of Anarchy with Dynamic Population via Stable Sequences%Price of Anarchy for Dynamic Games and Mechanisms
}\label{sec:poa}
\input{PoA-general}

\section{Stable Sequences via Greedy Algorithms}
\label{sec:greedy}
\input{matching}
    
\section{Stable Sequences via Differential Privacy}\label{sec:privacy}
\input{privacy}

\section{Applications of differential privacy in dynamic-population games}\label{ssec:privacy_applications}
 \input{privacy_applications}

%\begin{APPENDICES}
\setcounter{section}{0}
\def\thesection{\Alph{section}}
% \section{Proofs of Main Results}\label{sec:proofs}
% \input{omitted-proofs}

%\end{APPENDICES}

\subsection*{Acknowledgements}
We would like to thank Karthik Sridharan for pointing us to relevant adaptive regret learning literature and Janardhan Kulkarni for suggesting the direction of asymptotic optimality under participation uncertainty. Finally, part of the work was conducted when the authors were visiting the Simons Institute.

%% bibliography
\bibliographystyle{alpha}
\bibliography{bib1}

\section{Classical static-population smoothness proofs (Section~\ref{sec:preliminaries})}\label{app:prelims}
\input{app_prelims}

\section{Smoothness of first-price auctions with discrete bid spaces}
\label{sec:app-smoothness}
\input{app-smoothness}

\section{Asymptotic optimality with uncertain participation}
\label{app:asymptotic_uncertain_arrivals}
\input{large-market-smooth}

\end{document}

%% file: abstract.tex
%!TEX root = write-up.tex

We study the quality of outcomes in repeated games when the population of players is dynamically changing and participants use learning algorithms to adapt to the changing environment. Game theory classically considers Nash equilibria of one-shot games, while in practice many games are played repeatedly, and in such games players often use algorithmic tools to learn to play in the given environment. Most previous work on learning in repeated games assumes that the population playing the game is static over time.

We analyze the efficiency of repeated games in dynamically changing environments, motivated by application domains such as Internet ad-auctions and packet routing. We prove that, in many classes of games, if players choose their strategies in a way that guarantees low adaptive regret, then high social welfare is ensured, even under very frequent changes. In fact, in large markets learning players achieve asymptotically optimal social welfare despite high turnover. Previous work has only showed that high welfare is guaranteed for learning outcomes in static environments. Our work extends these results to more realistic settings when participation is drastically evolving over time.

%% file: introduction.tex
%!TEX root = main.tex
Quantifying the inefficiency caused by the selfish behavior of agents in systems of economic significance lies at the heart of algorithmic game theory. The \emph{Price of Anarchy} \cite{Koutsoupias1999} measures this inefficiency by comparing the welfare of the society where agents follow the instructions of a benevolent planner to the one where they behave in a selfish way. This notion has received great attention in both computer science and operations research, leading to a productive line of work in a plethora of application domains. For instance, routing games \cite{Roughgarden2002,Correa2003}, resource allocation \cite{Johari2004}, and online ad-auctions \cite{Caragiannis2015} have bounds on the efficiency loss due to incentives. This enables the platform operator to quantify the potential improvement in altering the system to better address incentives and make informed decisions on the trade-offs involved.

The recent emergence of repeated multi-player games has cast doubt on the aforementioned efficiency guarantees that rely on modeling selfish behavior via the equilibria of the one-shot game at each step. Unlike one-shot games, in settings such as online advertising, each individual advertising opportunity has negligible effect and their significance stems from the fact that these transactions occur repeatedly in very high frequency. Modeling the outcome of such interactions as the Nash equilibrium of the corresponding one-shot game is unrealistic for many reasons. Participants would need an unrealistic amount of information to know who else is participating at each moment, to compute a Nash equilibrium of the game. We note that the players repeatedly best responding to the current environment, doesn't lead to the Nash equilibria of the game even without the evolving participant list. The need for unrealistic amount of information is even more severe if we aim to model the outcome as an equilibrium of the repeated game.

To overcome the concerns about the Nash equilibrium notion, a more attractive model of player behavior is to assume that players use a form of algorithmic learning. In environments where the value or cost of a single interaction is low (a single ad opportunity or routing a single packet), using learning to find good strategies can be more attractive than precomputing effective strategies, and learning also enables the participants to adapt to changes in the environment. Blum et al \cite{BlumEL_PODC06, BlumHLR_STOC2008} suggest no-regret learning as a behavioral model of players, and study the outcome of such player behavior assuming that the exact same set of players engage repeatedly in the same game. The notion of no-regret learning assumes that players do at least as well as any single fixed strategy with hindsight. This assumption is guaranteed by many simple decentralized dynamics \cite{Hart2012,Freund1997} but crucially does not require
that players use a particular dynamic; it only posits that their behavior approximately satisfies the no-regret condition.  This is a broad behavioral assumption as it is satisfied by repeated occurrences of the one-shot Nash equilibrium outcomes, and even allows players to outperform the no-regret condition and do better than any single strategy; it only assumes that players avoid playing suboptimal actions at the existence of a consistently better alternative. As a result, it models well situations when participants use data from past interactions to adjust their strategies.
The no-regret assumption is supported by evidence in Bing ad-auction data by Nekipelov et al \cite{Nekipelov2015} using bidders with large budgets and frequent bid changes, who likely are using a form of learning to adjust their behavior.\footnote{Nisan and Noti \cite{NisanN16} study this assumption with human players, and their finding is more mixed. Bidders given low values, who may have been best off withdrawing from the game behaved irrationally, while players with higher assigned values did satisfy the assumption.}

On the efficiency front, most price of anarchy results, initially developed for Nash equilibria, extend to the broader learning behavioral assumption as long as the game repeated is  completely identical across rounds. This is thanks to extension theorems \cite{Roughgarden:2009:IRP:1536414.1536485,Syrgkanis2013} for a large class of games that satisfy the so-called \emph{smoothness property} and include routing games and simultaneous multi-item auctions. 

Although this learning-based analysis has improved the relevance of efficiency guarantees, there is still a major concern for the platform operator as the bounds only hold when players remain identical across all rounds of the game which rarely holds in applications of interest. In ad-auctions, advertisers may change their value for different keywords based on recent trends or marketing decisions. In packet routing, when a video conference ends, the configuration of data transmission alters. In transportation, when people switch employments or take vacations, similar changes in the routing patterns arise. Our work addresses this important consideration, highlighting the robustness of the learning-based analysis to changing environments and making price of anarchy guarantees relevant even when the population is constantly evolving. 

\subsection{Our contribution}
We provide a framework of \emph{dynamic-population repeated games}. Consider a game with $n$ players, such as ad-auctions with $n$ advertisers, or traffic routing with $n$ traffic streams. We 
refer to this as the stage game. The game 
is repeated each time step $t=1, 2,\ldots, T$, where after each step, each player may leave with some probability $p$ and be replaced with a different player. While we assume that players leave with a probability $p$ at random, the replacement player is worst case. This model of churn in the game is combining two important features: with  
a probability $p$ of departure, each player is expected to be in the game for $1/p$ time steps. We 
need to guarantee that $1/p$ is long enough that the player can learn to achieve small regret. At the same time, the worst case nature of the replacement of players enforces that learning behavior is necessary to perform well. For example, a routing choice may be great at the start, but due to newly arriving players may become congested as time goes on.

The model of learning behavior we use in this paper is that of adaptive learning, a form of learning well-suited for changing environments. Learning as a behavior has the great advantage that it can adjust to changes in the environment. Classical no-regret learning compares the player's outcome to a fixed strategy with hindsight. This is appropriate in games with short time spans, such as the data of \cite{Nekipelov2015,NisanN16} above. For repeated games played for longer periods, no-regret against a fixed strategy can be a very weak benchmark: it may be necessary for the learner to adjust her strategy due to
the changing environments. Natural extensions of no-regret learning can guarantee small regret also against a slowly changing sequence of strategies. We formally define this adaptive learning assumption and discuss algorithms that achieve this in Section \ref{ssec:learning_games}.

\paragraph{Our results.}
We provide a framework that transforms the static-population price of anarchy guarantees such as \cite{Roughgarden:2009:IRP:1536414.1536485, Syrgkanis2013} to corresponding \emph{dynamic-population} guarantees. Our guarantees are parametric and gracefully degrade with the amount of \emph{player turnover} but, surprisingly, can afford even a high rate of change with a significant fraction of the population turning over each step. Our framework relies on two properties for the underlying game, which are satisfied in multiple applications (such as simultaneous multi-item auctions and routing games). The first is that the game satisfies a variant of the smoothness property of \cite{Roughgarden:2009:IRP:1536414.1536485, Syrgkanis2013} which is crucial for the extension theorems even with static population. The second property is that the underlying optimization problem satisfies a stability condition: there exists a sequence of $\beta$-approximately optimal solutions that is not drastically changing; at a player turnover, the solution of \emph{few}\footnote{At a random player's arrival, at most a polylogarithmic number of other players is affected.} other players alters. As we explain below, we derive such stable approximately optimal solution sequences via two techniques and apply them to our applications of interest.
The first technique (Section~\ref{sec:greedy}) relies on greedy arguments, is related to the dynamic matching literature, and induces a decay in performance of $\beta=0.5$. The second technique (Section~\ref{sec:privacy}) draws a nice connection to the differential privacy literature and results to parametric degradation that tends to $\beta\rightarrow 1$ as the game becomes larger.  Our framework is described in detail in Section~\ref{sec:poa}.

Our main result is to prove an $\alpha\cdot \beta \cdot \gamma$ bound on the price of anarchy of the repeated game, assuming players' strategy choices approximately satisfy the dynamic no-regret condition, the stage game satisfies the smoothness assumption with price of anarchy $\alpha$ for the one-shot game, the underlying optimization problem has a $\beta$-optimal sequence of stable solutions, and $\gamma$ is a parameter allowing us to charge the player's regret error as a small fraction of the obtained social value, which gracefully degrades with the turnover probability, and goes to 1 as $p \rightarrow 0$. A dependence on $p$ is expected as, if the game is always abruptly and completely changing, we cannot expect to have stable solutions, and learning from the past cannot provide insightful guidelines for the future. 

Applying our framework to particular settings, we obtain guarantees for two important applications:
multi-item auctions and routing. 
Most of our results are about simultaneous multi-item auctions (such as first-price or second-price). We provide three results with different strengths.
\begin{itemize}
    \item Our first guarantee (Theorem~\ref{thm:matching_markets}) assumes that players are \emph{unit-demand} and provides a dynamic-population price of anarchy of $0.25(1-\eta)$ when the turnover probability is $p\lesssim \eta^3$. This bound stems from using the dynamic-matching stability approach leading to $\beta=0.5$, learning-based static-population price of anarchy guarantees of $\alpha=0.5$ (such as the one for simultaneous second-price auctions \cite{ChrKovSch16}) and can afford a turnover regret term of $\gamma\approx 1-\sqrt[3]{p}$ (up to logarithmic terms), allowing for close to a constant fraction of the population turnover each time step.
\item Our second guarantee (Theorem~\ref{thm:large_markets}) allows players to have more complex valuations satisfying the so-called \emph{gross substitutes} condition but requires each item to come with a large supply $B$; it provides a dynamic-population price of anarchy of $0.5(1-\eta)$ when the turnover probability is $p\lesssim \eta^5\frac{B}{n}$. This bound uses the same static-population price of anarchy guarantees of $\alpha=0.5$ but instead applies a differentially-private stability technique that leads to an effective $\beta=1-\eta$ only when the turnover term is $\gamma\approx 1-\eta-\sqrt{pn/B}$. As a result, the guarantee gives improvement when the market is thick, e.g., $B=\Theta(n)$ but can be meaningless in small-supply settings where the first guarantee still applies.
\item Our third guarantee (Theorem~\ref{thm:large_markets_uncertainty}) provides asymptotic optimality $1-\eta$ (when the turnover probability is $p\lesssim \eta^5\frac{B}{n}$) and when there exists some uncertainty in the arrival of each player at every round (each player does not show up with probability $q$ even if they are in the game). With this additional assumption, we can use enhanced static-population guarantees of $\alpha\approx 1-\frac{1}{\sqrt{q(1-q)B}}$ \cite{FeldmanILRS_STOC16} while keeping the parameters $\beta$ and $\gamma$ as before. As a result, when the market is sufficiently thick, the price of anarchy goes to $1$ as the turnover probability $p$ becomes small.
\end{itemize}

The second application is \emph{atomic unsplittable routing}; since this is a cost-minimization setting, all the approximation factors are above $1$. For this class, the learning-based analysis \cite{Roughgarden:2009:IRP:1536414.1536485} provides static-population price of anarchy guarantees (e.g. $\alpha=2.5$ for affine latency functions \cite{Christodoulou:2005:PAF:1060590.1060600}). 
Our guarantee (Theorem~\ref{thm:routing}) extends this to a dynamic-population price of anarchy of $2.5+\eta$ when $p\lesssim \eta^4/m^{10}\polylog(n,m)$ where $m$ is the number of edges in the network, when the number of players is relatively large. Note that again this turnover probability depends only logarithmically on the number of players so allows for a nearly constant turnover when the number of players is relatively large. 

\paragraph{Our techniques:}
To achieve the claimed welfare guarantees we assume that all players are learning and take advantage of the power of no-regret learning. Classical no-regret learning guarantees that the learner achieves utility as good as any \emph{fixed} strategy with hindsight. With the changing participants of a dynamic population game, the competitors over time can change drastically, and participants need to adjust to the changing environment; a fixed strategy with hindsight is therefore not strong enough to ensure good outcomes. However, learning can do better: rather than a fixed single strategy as a benchmark, \emph{shifting regret} can guarantee to do well against an occasionally changing benchmark \cite{Herbster1998,DBLP:journals/corr/LuoS15}. As one would expect, the effectiveness of learning degrades with the number of changes in the benchmark.

In proving welfare guarantees for learning in repeated games \emph{with fixed and unchanging} set of participants, traditional learning-based smoothness approaches \cite{RoughgardenJACM,Syrgkanis2013} rely on the fact that the set of players is static. In this case, the optimal solution is the same at each iteration and the strategy $s^\star_i$ associated with player $i$'s solution in this fixed optimal solution is therefore a good fixed benchmark; hence classical no-regret learning achieves at least as good performance.   

In a dynamic population setting, the optimal solution changes as the player set evolves. To guarantee high welfare in these evolving games, one idea would be to use the fact that, by the strengthened \emph{shifting learning} assumption 
players have no-regret against a shifting sequence of $s_i^{\star,t}$ associated with the optimal solutions $\opt^{t}$ at each time $t$. To understand whether this is
effective, we need to investigate \emph{how fast does this benchmark actually change as a function  
of the turnover probability?} Unfortunately, the sequence of optimal solutions can be highly unstable for any player. Every time that a player turns over, the new optimal solution can change drastically (e.g., for an assignment problem the change is along an augmenting path that can be arbitrarily long). With frequent changes in the benchmark, learning is ineffective (the learning error dominates), so for learning to be effective against the benchmark of the optimal solution, one would need to require that new players arrive extremely rarely.

To make the price of anarchy guarantees robust for a high turnover, we need to avoid comparing to the unstable optimal solution sequence and instead use a \emph{very stable} benchmark, i.e. that does not drastically alter with a type change of a single player. We offer two techniques to show that many optimization problems have close-to-optimal stable solutions that can be used as benchmarks. In Section \ref{sec:greedy} we show that for the assignment problem, the greedy algorithm can be adapted to result in a stable solution.  Using this greedy solution as a benchmark
loses a factor of $\beta=0.5$ in efficiency compared to the optimal allocation, however, this benchmark allows for a close-to-constant fraction of the population to turn over every single round ($p$ only depends logarithmically on the number of players) and still guarantee high social welfare.

In Section \ref{sec:privacy} we show an interesting connection between stable solution sequences and differential privacy, allowing guarantees where the extra decay $\beta$ caused by the requirement of using a stable benchmark parametrically degrades with some notion of \emph{largeness} of the game. Joint differentially private algorithms \cite{DBLP:conf/innovations/KearnsPRU14} guarantee that, if one player's input changes, the probability that the output for other players changes is small. This is in accordance with our stability requirement -- the expected number of changes that a single turnover causes is small (since all players are affected with a small probability). We can therefore use the existence of such joint differentially private algorithms for multi-item allocation problems \cite{hsu2016private} and online routing \cite{DBLP:journals/corr/RogersRUW15} to derive efficiency guarantees for the corresponding games that tend to the price of anarchy $\alpha$ of the static-population game as the game becomes larger. Crucially, the results again allow for a close-to-constant fraction of the population to turn over every single round.

We should note that the benchmark solutions discussed here are never explicitly computed and the players do not need to know anything about them or about the changes occurring in the game --  the players just use their learning algorithms. The existence of such \emph{stable solution sequences} implies that learning players have no regret against them and this allows to extend the efficiency guarantees to dynamic population settings even with a large churn in the player set.

\Xomit{OLD VERSION THAT IS MORE TECHNICAL
To demonstrate the technical contribution of the paper, we consider the learning-based analysis in a very simple game: simultaneous second-price auctions where $n$ players have value $0$ or $1$ for $m$ items and are \emph{unit-demand}, i.e. have no value for more than one item. Consider the optimal solution (disregarding incentives) for this setting -- this is a maximum matching $\emph{M}^{\star}$. One of the strategies for player $i$ is to bid their true value at the item they get allocated in $\emph{M}^{\star}$ (if such an item exists) and bid $0$ in all others. Since the player is a no-regret learner, the utility she is getting is close to the utility she would have got had she played this fixed strategy $s_i^{\star}$ across rounds; she suffers a small amount of regret $\mathcal{O}(\sqrt{T})$ that, in average, vanishes as the time horizon $T$ goes to infinity. The smoothness property establishes that the sum of these unilateral deviations of all players produces a decay of at most $\alpha=0.5$ compared to the optimal solution. Combining the no-regret and smoothness properties, the learning-based analysis establishes that, in static population, the average social welfare with selfish players is at least an $\alpha$ \etcomment{you are liberally using $\alpha, \beta$ and $\gamma$ in this section, seem to assume that the reader memorized what they stood for. I would suggest we should give them names, and use both that and the Greek letter. Not sure what are OK names, but $\alpha$ can be the classical PoA loss, $\beta$ the stability loss, and $\gamma$ the learning loss.} fraction of the optimal social welfare if we disregard incentives minus a term $\nicefrac{n}{\sqrt{T}}$ that vanishes over time $T$. Crucially for the analysis, $s_i^{\star}$ is the same for all rounds as the no-regret property guarantees that players have \emph{no regret} against deviations to a \emph{fixed} strategy.

In a dynamic population setting, the optimal solution changes as the types of the players evolve, which causes a first roadblock in extending the results since players need to have no-regret against a shifting sequence of $s_i^{\star,t}$ associated with the optimal solution $\emph{M}^{\star,t}$. Fortunately, there exist algorithms that achieve small \emph{shifting regret} against a changing benchmark \tlcomment{add references}. Typically the regret guarantees degrade with the number of changes in the benchmark $K_i$ as a function $\mathcal{O}(\sqrt{K_i\cdot T})$ (c.f. related work for a more elaborate discussion). However, this term no longer vanishes across time if the number of changes is relatively large and could dominate the social welfare making the results vacuous. Therefore, to obtain meaningful efficiency guarantees, we need a bound on the effect of this term which is captured by our parameter $\gamma$ and is related to how large the turnover is. Hence a crucial question is: \emph{How fast does this benchmark actually change as a function  
of the turnover probability?} Unfortunately, the sequence of optimal solutions can be highly unstable for any player. Every time that a player turns over, the new optimal solution is an augmenting path that can alter the solution of all the players.\footnote{In fact, the adversarial nature of incoming types allows it to affect at least half the players at a random arrival.} This means that the allocation of player $i$ is affected $K_i\sim n\cdot p\cdot T$ times (at every round, the probability of change is the probability that \emph{any} player's type turns over). This is clearly too high and results to a dependence $\gamma\sim 1-\sqrt{n\cdot p}$ which becomes constant $1-\epsilon$ only when $p\sim \nicefrac{\eps^2}{n}$ allowing for changes in the population to happen only sporadically, and is clearly not satisfying for the applications we discussed above.

For robustness against higher turnover, we need a \emph{more stable} benchmark, i.e. that does not drastically alter with a type change of a single player. Consider instead the allocation induced by the greedy algorithm that maximally assigns items to players, favoring pairs that existed in the allocation of the previous round. Under this scenario, a player departure does not affect the allocation of any player previously assigned to an item and can only cause change in their own allocation and possibly one more player that can be matched to the item that she leaves unassigned. \etcomment{one more player? Actually, doesn't it have a rippling effect of many players downstream? its just that the potential function argument proves that this is small on average.} Therefore the number of changes that an average player observes in this example is at most $K_i\sim 2\cdot p\cdot T$, leading to a parameter $\gamma\sim 1-\sqrt{2\cdot p}$ that allows a constant fraction of the population to change every single round ($p\sim \epsilon^2$) while retaining efficiency guarantees of $\gamma=1-\epsilon$. The downside is that the induced allocation is not optimal but rather within $\beta=0.5$ of the optimal solution which is the root of the extra drop in efficiency in our applications using this technique. To generalize this technique beyond $\crl{0,1}$ values, we introduce an elegant potential function argument that allows the guarantees in Section \ref{sec:greedy}. We should note that the allocations discussed are never explicitly computed and the players do not need to know anything about them or about the changes occurring in the game-- the game is simultaneous second-price auction and the players just use their learning algorithm. The existence of such \emph{stable solution sequences} implies that learning players have no regret against them and this allows to extend the efficiency guarantees to dynamic population settings even with a large churn in the player set. 

Exploring the generality of the above framework, we subsequently show an interesting connection between stable solution sequences and differential privacy, allowing guarantees where the extra decay $\beta$ parametrically degrades with some notion of \emph{largeness} of the game. Joint differentially private algorithms \cite{DBLP:conf/innovations/KearnsPRU14} guarantee that, if one player's input changes, the probability that the output for other players changes is small. This is in accordance with our stability requirement -- the expected number of changes that a single turnover causes is small (since all players are affected with a small probability). We can therefore use the existence of such joint differentially private algorithms for online routing \cite{DBLP:journals/corr/RogersRUW15} and multi-item allocation problems \cite{hsu2016private} to derive efficiency guarantees for the corresponding games that tend to the price of anarchy $\alpha$ of the static-population game as the game becomes larger. Crucially, the results again allow for a constant fraction of the population to turn over every single round ($p$ is independent of the number of players up to logarithms) and therefore captures the settings of interest.
}

\subsection{Related work}
Our work lies in the interplay of learning and game theory in dynamic environments. In what follows, we briefly discuss relevant related work in each area, and clarify the key differences.

\paragraph{Dynamic games.}
Dynamic games have a long history in economics, dynamical systems and operations research, see for example the survey books of Başar and Oldser \cite{Basar1998} and Van Long \cite{VanLong2010}. The classic approach of analyzing behavior in such dynamic games is to assume that players have prior beliefs about their competitors and that their behavior will constitute a \emph{perfect Bayesian equilibrium} or  refinements of it such as the \emph{sequential equilibrium} of Fudenberg and Tirole \cite{Fudenberg1991} or \emph{Markov perfect equilibrium} of Maskin and Tirole \cite{Maskin2001}. The informational and rationality assumptions that such equilibrium solution concepts impose on the players and their computational intractability casts doubt on whether players in practice and in complex game-theoretic environments such as packet routing or internet ad-auctions would behave as prescribed by such an equilibrium behavior.

Equilibrium-like behavior might be more plausible in large game approximations (see e.g. the work of Kalai and Shmaya \cite{Kalai2015}). A natural approximation to equilibrium behavior in large game situations that has been recently extensively analyzed in economics  and in operations research (and particularly, in auction settings) is that of the mean field equilibrium \cite{Weintraub2008, Adlakha2013, Iyer2014, Adlakha2015, Balseiro2015} and relaxations thereof, such as oblivious equilibrium \cite{Weintraub2006, weintraub2010computational}.  These advances bring equilibrium theory much closer to practice and make more realistic assumptions than tradition sequential equilibrium theory. Moreover, they allow for long-range dependencies in the player's learning (such as budget constraints), which we do not consider here (see also relevant discussion in subsequent paragraphs on learning in games). However, even these large game approximations require the players to form almost correct beliefs about the competition and to exactly best-respond to these approximate large-market beliefs. Moreover, the approach requires that the environment either is stochastically stable or evolves in a known stochastic manner. On the contrary our dynamic-population model allows for adversarial changes and our analysis attempts to analyze even constantly evolving and never converging behavior. Moreover, our assumption that players satisfy a form of no-regret condition for adaptive learning does not impose that players possess or form any beliefs on the competition. Most of the algorithms that achieve this guarantee only require that the player is able to see the utility that each of her strategic options would have given in-retrospect, in past time-steps. Furthermore, our approach also applies to small markets although our results improve in the large-market regime. Finally, we note that in the large market regime, prior work of \cite{BalseiroGur2019}, shows a strong connection between no-regret learning and mean field equilibria, even when there are budget constraints (albeit in a stationary world). It is an interesting avenue for future work, whether such connections extend to the non-stationary setting we consider here.

There is a large literature on truthful mechanisms in a dynamic setting analogous to our dynamic player population model, where the goal is to truthfully implement a desired outcome with dynamically changing populations of users with private value. This line of work goes back to \cite{ParkesS_NIPS03} in the computer science literature, but has been also considered much earlier with queuing models by \cite{Dolan78}. In subsequent work, Cavallo et al. \cite{CavalloPS2014} offer a generalized VCG mechanism in an environment very similar to the one we are considering with departures and arrivals, and also provides a nice overview of work in truthful mechanisms in a dynamic setting; the reader is also referred to the survey on dynamic auctions by \cite{bergemann-said-2010}. More recently, there has been significant work in understanding dynamic mechanism design in more complicated settings such as not knowing the buyer distributions in advance \cite{MirrokniPaesLemeTAngZuo18} or budget constraints \cite{BalseiroBesbesWein19}. In contrast to these works, we do not design the mechanism but rather analyze the outcomes of fixed broadly used mechanisms such as first and second price auctions, or other games such as routing. 

\paragraph{Learning in games.} In recent years, learning received attention as a behavioral assumption in games  especially with the rise of online advertising which, as discussed before, makes its deployment widespread and effective. There is a large literature analyzing learning in games, dating back to the work on fictitious play by Brown and Robinson \cite{brown_51,Robinson51aniterative}. For an overview of this area, the reader is referred to the books of Fudenberg and Levine \cite{FudenbergLevine} and Cesa-Bianchi and Lugosi \cite{CesaBianchi2006}. A standard notion of learning in games is that of no-regret learning. The notion of no-regret against the best fixed action in hindsight dates back to the work of Hannan \cite{H-57}, and is also referred as \emph{Hannan consistency}. There are many learning algorithms achieving this guarantee such as regret matching \cite{hart00} and multiplicative weights updates \cite{Freund1997}.
In the context of learning to bid in repeated auctions, these learning algorithms work well when players are bidding only on a few items each round (even if there may be many possible items to bid on), but finding good learning strategies is computationally hard when applied to general multi-item auctions \cite{CaiPapadimitriou2014,DaskalakisSyrgkanis2016}. In our auction results, we assume that there is a limit in the number of items that any player can bid on which makes learning effective.

Most efficiency guarantees originally developed for Nash equilibria extend to static-population games via the smoothness framework \cite{RoughgardenJACM,Syrgkanis2013}. Our work extends this to more realistic settings that can be drastically evolving over time.

Another line of work involves mechanism design questions when the participants in the setting use no-regret learning algorithms \cite{BravermanMaoSchneiderWeinberg18,DengSchneiderSivan19}; these papers suggest that classical no-regret learning against a fixed benchmark can be easily attacked by revenue-maximizing auctioneers, further supporting the need for more sophisticated learning algorithms such as the ones we study in this work.

\paragraph{Learning in dynamic games.}
The notion of no-regret learning against time-varying benchmarks, as opposed to fixed actions, traces back to Herbster and Warmuth \cite{Herbster1998} who provided guarantees compared to the best sequence of $k$ experts where $k$ is known in advance. The stronger notion of adaptive regret which we consider in this work was formalized by Hazan and Seshadhri \cite{HazanS07} and near-optimal adaptive regret guarantees were achieved through a series of algorithms \cite{Lehrer2003,Blum2007, Cesabianchi2012,DBLP:journals/corr/LuoS15} (we further discuss this in Section \ref{ssec:learning_games}). One important trait of these algorithms is that they display some sort of recency bias, in the sense that the influence of past steps decays as time goes by; this is also supported by experimental evidence \cite{Fudenberg2014}, which suggests that humans display such forms of recency bias when making repeated decisions.

Finally, subsequent to the preliminary version of this paper, Foster et al. \cite{conf/nips/FosterLLST16} showed that the efficiency guarantees we present in this paper hold under a weaker behavioral assumption. The \emph{adaptive learning} assumption we analyze in this paper is one way to achieve low regret against an adaptive benchmark with unknown number of changes in the comparator sequence. Foster et al. \cite{conf/nips/FosterLLST16} show that a multiplicative relaxation of the regret notion similar to the one used by Blum and Mansour \cite{Blum2007}, suffices for the efficiency guarantees we prove, and is satisfied by minor modifications of most no-regret algorithms. The latter strengthens the results of this paper by weakening the required behavioral assumption since our efficiency guarantees hold even if players can select algorithms from a broader class. In the remainder of the paper, we use the original \emph{adaptive learning} behavioral assumption and refer the reader to \cite{conf/nips/FosterLLST16} for further discussion on the weaker assumption. 

%% file: preliminaries.tex
%!TEX root = write-up.tex
In this section, we formalize the model we consider, provide background from learning theory and game theory that will be essential in our results, and discuss the applications we consider.

\subsection{Model}
\paragraph{Games and mechanisms.}
We consider game settings played repeatedly, where the population of players is evolving over time. Let $G$ be an $n$-player normal form \emph{stage game} and assume that game $G$ is played repeatedly for $T$ rounds.\footnote{Although the game rules remain unaltered across rounds, the changes in participants affect the payoff matrix.} Each player $i$ has a strategy space $S_i$, with $\max_i|S_i|=N$, a type $v_i\in \V_i$ and a cost function $c_i(s; v_i)$ that depends on the strategy profile $s \in \times_i S_i$, and on her type. We denote with $C(s;\vec{v})=\sum_{i\in [n]}c_i(s;v_i)$ the social cost, where $s$ is a strategy profile and $\vec{v}$ a type profile. We also analyze the case when the stage game is a utility maximization mechanism $M$, which takes as input a strategy profile and outputs an allocation $X_i(s)$ for each player and a payment $P_i(s)$. We assume that players have quasi-linear utility $u_i(s;v_i)=v_i(X_i(s))-P_i(s)$ and the welfare is the sum of valuations (sum of utilities of bidders and revenue of auctioneer): $W(s;\vec{v})=\sum_{i\in [n]} v_i(X_i(s))$.

In all the settings we study, one can define the optimal solution of the underlying optimization problem. Let $\X^n$ be the ``feasible solution space'' of the setting disregarding incentives. In network congestion games this corresponds to the set of feasible integral flows, while in combinatorial auctions it is the set of feasible partitions of items to bidders.
We overload the social cost and welfare notations and, for a feasible solution (or allocation) $\vec{x}\in \X^n$, use $C(\vec{x};\vec{v})$ and $W(\vec{x};\vec{v})$ to denote the social cost or welfare of the solution\footnote{Overloading the notation does not create ambiguity between $C(x,v)$ and $C(s,v)$, assuming the strategy sets $S_i$ are disjoint from the possible solutions $\X$.}. We denote the optimal social cost or welfare for a type profile $\vec{v}$, as $\opt(\vec{v})=\min_{\vec{x}\in \X^n} C(\vec{x};\vec{v})$ and $\opt(\vec{v})=\max_{\vec{x}\in \X^n}W(\vec{x}; \vec{v})$ respectively.

\paragraph{Repeated games and mechanisms with dynamic population.} We introduce the study of repeated game settings when the population of the player evolves over time. Our model is formalized in the following definition.
\vspace{0.1in}
\begin{definition}[Repeated game/mechanism with dynamic population] A \textbf{repeated game with dynamic population} consists of a stage game $G$ played for $T$ rounds. Let $P^t$ denote the set of players at round $t$, where each player $i\in P^t$ has a private type $v_i^t$. After each round, every player independently exits the game with a (small) probability $p>0$ and is replaced by a new player with an arbitrary type. The utility of players is additive across rounds. This repeated game is denoted by $\Gamma=(G,T,p)$; similarly $\M=(M,T,p)$ denotes a corresponding mechanism. 
\end{definition}

Our model of dynamic population assumes that after each round every player independently exits the game with a turnover probability $p>0$; each player is expected to participate in $1/p$ rounds. To keep our model simple, we assume that when a player exits, she is replaced by a new participant. This assumption guarantees that there are exactly $n$ players in each iteration, with a $p$ fraction of the population changing each round in expectation. We make no assumption on the types of the new arriving players which can be selected adversarially. Most of our results can be extended assuming only an analogous bound on the expected number of departures,  allowing departures to be correlated or even chosen adversarially. 

To simplify notation, we use \emph{player $i$} to denote the current $i$-th player, where this player is replaced by a new $i$-th player with probability $p$ each round. An alternate view of the dynamic player population is to think of players as changing types after each iteration with a small probability $p$. We refer to such a change as \emph{player $i$ switches} or \emph{turns over}.

\paragraph{Basic notation.} For any quantity $x$ we denote with $x^{1:T}$ the sequence $x^1,\ldots, x^T$. For instance, $v_i^{1:T}$ denote the sequence of types of player $i$ produced by the random choice of departing players and the choices of the adversary regarding arriving players.

\subsection{Background from learning theory and game theory}
\label{ssec:learning_games}
\paragraph{Adaptive Learning in Dynamic Environments.} 
No-regret learning provides guidance on how to decide in an online manner across a set of alternatives. Consider an arbitrary loss vector across the possible actions. For a cost-game, the loss of an action corresponds to the cost the player incurs if they follow this action; for a utility game, the loss is the difference between the maximum possible utility and the player's utility. Consider a player who has $N$ possible actions. In defining regret and no-regret learning, we focus on a single player and hence temporarily drop the index $i$ for the player from the notation. We use $\ell(s,t)$ to denote the loss (cost or lost  utility) of the player if she plays strategy $s$ at round $t$. 
We assume without loss of generality that $\ell(s,t)$ is a value in $[0,1]$, but make no further assumption about the sequence of losses. Note that even with a stable set of players the value $\ell(s,t)$ may vary over time, depending on the strategies chosen by other players. A player achieves no-regret if she does at least as well over a period of time as the best choice $s^{\star}$ with hindsight.
\vspace{0.4em}
\begin{definition}[Regret]
The regret of a strategy sequence $s^{1:T}$ is defined as:
$$
R(1,T)=\max_{s^{\star}}\sum_{t=1}^T \prn*{\ell(s^t,t) - \ell(s^{\star},t)}
$$
\end{definition}
There are many simple algorithms (for example, see the book of Cesa-Bianchi and Lugosi \cite{CesaBianchi2006}) that achieve regret $O(\sqrt{T\log(N)})$ against any (adversarial) sequence of loss values $\ell(s,t)$.

In dealing with changing environments, we require a stronger assumption on the learning of the players, since the players need to adapt their strategies to the environment. We use the notion of adaptive regret introduced by Hazan and Seshadhri \cite{HazanS07} which captures the regret over (long) time intervals $[\tau_1,\tau_2)$, in addition to the regret of the whole sequence.

\vspace{0.1in}
\begin{definition}[Adaptive Regret]
The adaptive regret of strategy sequence $s^{1:T}$ in the time interval $[\tau_1,\tau_2)$ is defined as:
\begin{equation*}
R(\tau_1,\tau_2)=\max_{s^{\star}}
\sum_{t=\tau_1}^{\tau_2-1}\prn*{\ell(s^t,t)-\ell(s^{\star},t)}
 \end{equation*}
\end{definition}

We assume that all players use learning algorithms that achieve low adaptive regret. For simplicity of presentation we assume that, for a constant $C_R$, $\mathbb{E}[R(\tau_1,\tau_2)]\leq C_R\sqrt{(\tau_2-\tau_1)\ln(N\tau_2)}$ and refer to this assumption as \emph{``The players use adaptive learning algorithms with constant $C_R$}''.\footnote{Our results  smoothly degrade if we instead assume that players achieve adaptive regret that is some other sublinear concave function of the interval's length $(\tau_2-\tau_1)$.} This assumption can be achieved by the algorithms of Luo and Schapire \cite{DBLP:journals/corr/LuoS15}, or Daniely, Gonen, and Schalev-Shwartz \cite{pmlr-v37-daniely15}. In fact, any algorithm that achieves a \emph{sleeping experts} regret guarantee \cite{blum1997empirical,freund1997using,Blum2007} automatically satisfies this assumption as noted by Luo and Schapire \cite{DBLP:journals/corr/LuoS15}. For simplicity of presentation, we use $C_R=1$ and refer to the behavioral assumption as \emph{``The players use adaptive learning algorithms''}

\paragraph{Solution-based Smoothness in Games and Mechanisms.}
Smooth games were introduced by Roughgarden \cite{Roughgarden:2009:IRP:1536414.1536485} as a general framework bounding the price of anarchy in games. He also showed that smoothness based price of anarchy bounds extend to outcomes in repeated games where the set of players is fixed throughout the period and all players use no-regret learning.

We require a somewhat more general variant of smooth games, that compares the cost or utility resulting from a strategy choice to the social welfare of a specific solution, rather than comparing to the social optimum. For two strategy vectors $\vec{s}$ and $\vec{s^{\star}}$ we use $(s_i^{\star},s_{-i})$ to denote the vector where player $i$ uses strategy $s_i^{\star}$ and all other players $j$ use their strategy $s_j$.
\vspace{0.1in}
\begin{definition}[Solution-based smooth game]
A cost-minimization game $G$ is solution-based $(\lambda,\mu)$-smooth for $\lambda>0$ and $\mu<1$, if for any type profile $\vec{v}$, for any feasible solution $\vec{x}\in\X^n$ and any player $i$, there exists a strategy $s^{\star}_i\in S_i$ depending only on her type $v_i$ and her part of the solution $x_i$ such that for any strategy profile $s$
\begin{equation*}\label{eqn:target}
\sum_i c_i(s_i^{\star}(v_i,x_i),s_{-i};v_i) \leq \lambda C(\vec{x};\vec{v}) + \mu C(\vec{s};\vec{v})
\end{equation*}
\end{definition}
The definition of smoothness in games used in the conference paper \cite{Roughgarden:2009:IRP:1536414.1536485} required that for any pair of strategies $\vec{s}$ and $\vec{s^{\star}}$ we have $\sum_i c_i(s^{\star}_i,s_{-i};v_i) \leq \lambda C(\vec{s}^{\star};\vec{v}) + \mu C(\vec{s};\vec{v})$. This definition turned out to be too strong, and for example it fails in the important application of Generalized Second Prize Auction (GSP) \cite{PaesLemeT10}. The definition used in the journal version \cite{RoughgardenJACM} and since then, requires only the existence of a strategy vector $\vec{s}^{\star}$ satisfying $\sum_i c_i(s^{\star}_i,s_{-i};v_i) \leq \lambda \opt(\vec{v}) + \mu C(\vec{s};\vec{v})$, effectively requiring the above condition only for a strategy vector $\vec{s}^\star$ associated with the the optimal solution $\vec{x}^\star$. Our definition above is between the two: requires the condition to hold for a strategy vector $\vec{s}^\star(\vec{v};\vec{x})$ associated with any solution $x$, not only the optimal one. Note that, when $\vec{x}$ is the optimal solution, we recover the traditional examples of smooth games (including GSP), as the deviating strategy $s^{\star}$ usually depends on other players' types through her part of the optimal solution $x_i^{\star}(\vec{v})$.

A game that is   solution-based $(\lambda,\mu)$-smooth then it is also  $(\lambda,\mu)$-smooth in the sense of \cite{Roughgarden:2009:IRP:1536414.1536485} (using solution-based smoothness for the optimal solution $\vec{x}^{\star}(\vec{v})$), and the game has price of anarchy bounded by $\lambda/(1-\mu)$, and the average social cost of no-regret learning outcomes is also bounded by $\lambda/(1-\mu)\opt$. More generally, for static-population games:
\vspace{0.4em}
\begin{proposition}[slightly extending \cite{RoughgardenJACM}]\label{prop:smooth_game} If a game is solution-based $(\lambda,\mu)$-smooth 
and players have
regret at most $\sqrt{T}$ for the strategy $s^\star_i(v_i,x_i)$ corresponding to the solution $x\in\X^n$, the average expected cost is at most $\frac{\lambda}{1-\mu}\prn*{C(\vec{x};\vec{v})+\frac{n}{\sqrt{T}}}$.
\end{proposition}
\vspace{-0.4em}
To familiarize the reader with the smoothness analysis, we provide the proof of both this proposition and the next one (Proposition~\ref{prop:smooth_mechanism}) in Appendix \ref{app:prelims}.

Observe that, when $T\geq \prn*{\frac{n}{\eps\cdot C(\vec{x};\vec{v})}}^2$, the above bound becomes $\frac{\lambda(1+\epsilon)}{1-\mu}\cdot C(\vec{x};\vec{v})
$.  Note also that it was crucial that strategy $s_i^{\star}$ was fixed across rounds since we used that player $i$ does not regret not deviating to it. Since $s_i^{\star}$ is a function of the underlying optimization problem, this is no longer the case in an evolving population as departures of players change the underlying optimization problem and may affect $s_i^{\star}$. In Section \ref{sec:poa}, we show how the stronger behavioral assumption of adaptive learning can help extend the aforementioned analysis.

The main result of our paper
is about mechanisms. For mechanisms we use the version of the smoothness definition of Syrgkanis and Tardos \cite{Syrgkanis2013} which assumes that players have quasi-linear utilities.
We again define a mechanism to be solution-based smooth and allow the choice of strategy $s^{\star}$ to depend only on the player's part of the solution $x_i$ and her type $v_i$. More formally:
\vspace{0.4em}
\begin{definition}[Solution-based smooth mechanism]
A mechanism $\mathcal{M}$ is solution-based $(\lambda,\mu)$-smooth for $\lambda,\mu\geq 0$, if for any valuation profile $\vec{v}$, any solution $x \in \X^n$, and any player $i$, there exists a deviating strategy $s_i^*\in S_i$ depending only on $v_i$ and $x_i$ such that for any strategy profile $s$,
\begin{equation*}
\sum_i u_i(s_i^*(v_i,x_i),s_{-i};v_i) \geq \lambda W(\vec{x};\vec{v}) - \mu\rev(s).
\end{equation*}
where $\rev(s) = \sum_{i=1}^{n} P_i(s)$. 
\end{definition}
Syrgkanis and Tardos \cite{Syrgkanis2013} proved that a $(\lambda,\mu)$-smooth mechanism has price of anarchy bounded by $\max(\mu,1)/\lambda$, and the average social welfare of no-regret learning outcome is also at least $(\lambda/\max(\mu,1)) \opt(\vec{v})$. Analogously, for static-population mechanisms, we obtain:

\begin{proposition}[slightly extending \cite{Syrgkanis2013}]\label{prop:smooth_mechanism} 
\vspace{0.1in}If a mechanism is solution-based $(\lambda,\mu)$-smooth 
players have
 regret at most $\sqrt{T}$ for a strategy $s^\star_i(x_i,v_i)$ for a solution $x\in \X^n$, then the average expected social welfare is at least $\frac{\lambda}{\max(\mu,1)}\prn*{W(\vec{x};\vec{v})-\frac{n}{\sqrt{T}}}$.
\end{proposition}

\subsection{Applications}\label{ssec:applications}
We consider a welfare-maximization mechanism (simultaneous multi-item auctions) and a  cost-minimization game (atomic congestion game) as the main application of our techniques.

\paragraph{Simultaneous multi-item auctions.}
The auction settings we consider include repeated auctions of $\widetilde{m}$ distinct item types, each of which appears with a limited supply $B$ (of identical copies); the total number of items is therefore $m=\widetilde{m}\cdot B$. In Section~\ref{sec:greedy}, we assume that the supply is $B=1$ (identical copies are treated as distinct items) while in Section~\ref{sec:privacy}, we assume that we have access to a larger supply $B$. In all these settings, each player submits bids for each item type and the highest bidder receives the item (ties are broken arbitrarily). The settings differ by the payment that this player ends up paying: in first price-auctions she pays her bid, in threshold-price auctions she pays the minimum price that would have her win the item (second bid when $B=1$ and more generally, $(B+1)$-th bid for general $B$), while in hybrid auctions she pays a combination of the two. The results of Section~\ref{sec:greedy} apply to any of these settings and are provided for simplicity for first-price auctions, while the results of Sections~\ref{sec:privacy} and \ref{ssec:privacy_applications} require smoothness results that only hold for threshold-price auctions so only apply in this case. For our asymptotically optimal efficiency (Section~\ref{ssec:asymptotic_uncertainty}), we also assume some uncertainty in participation of each existing player at every round similar to the work of Feldman et al. \cite{FeldmanILRS_STOC16}. 

Each player is assumed to have a valuation function for each set of items she obtains. Valuations are additive across time which models perishable items such as advertising opportunity where a player aims to repeatedly win items in each period she participates. For each round $t$, we use $v_i^t(A)$ to denote the valuation of the $i$-th player if she obtains set $A\subset [m]$. For all our results, we assume that the valuations are non-negative and normalized to be at most $1$; we also assume that values are expressed as multiple of $\rho>0$; in particular, the minimum non-zero value for an item is $\rho$. In Section~\ref{sec:greedy}, we assume that the players are unit-demand (their value is equal to the maximum value across items they obtained). In Section~\ref{sec:privacy} we consider players in large markets with more complex valuations (satisfying the gross substitute property, formally defined in that section).

Finally, since the learnability of the setting depends on the number of strategies available to the player, we restrict the players to bid to at most $d$ item types at each round and discretize their bids to multiples of $\zeta\rho$ for a small constant $\zeta>0$. As a result, the number of strategies available to the learner is equal to $N=\binom{m}{d}\left(\frac{1}{\zeta\rho}\right)^d\leq \left(\frac{m}{\zeta\rho}\right)^d$. For ease of presentation, in Section~\ref{sec:greedy}, we assume that $d=1$.

\paragraph{Atomic congestion game.}
In the atomic congestion game, we assume that we have a set of congestible elements $E$ (and let $m=|E|$), each element $e$ has a latency function $\ell_e(x)$, or cost, that is monotone non-decreasing in the congestion $x$. Given some selection of sets $s_i\subseteq E$ for each player $i$, the congestion in an element $e$ is the number of players that have selected it: $x_e(s)=|\{i:e\in s_i\}|$, and the cost of player $i$ is then the sum $\sum_{e\in s_i} \ell_e (x_e(s))$.

A player's type $v_i^t$ denotes the possible subsets of the element set she can select. For example, in the routing game on a graph, the type of a player $i$ is a source-sink pair $(o_i,d_i)$, and her strategy
is the choice of a path from $o_i$ to $d_i$ in the graph. We assume that a player's cost is infinity if her solution is not one of the selected sets. Thus the number of strategies available to the player is the number of $(o_i,d_i)$, paths in the graph and thereby $N$ is the maximum number of such possible paths across possible source-sink pairs.

%% file: PoA-general.tex
%!TEX root=main.tex
In this section, we provide our general efficiency theorems both for games and mechanisms with dynamic population. Specifically, we formalize the connection between adaptive learning, solution-based smoothness of the game, and the existence of \emph{approximately-optimal} and \emph{stable} sequences for the underlying optimization problem. The efficiency theorems serve as meta-theorems that reduce the quest for efficiency guarantees with dynamic population to the existence of approximately optimal stable solution sequences. Note that such stable sequences just need to exist, are only used in the proofs, they do not need to be known by the players, or even be constructed. In the following sections, we instantiate this meta-theorem to specific applications by showing how one can derive such stable approximately optimal sequences from greedy algorithms (Section~\ref{sec:greedy}) as well as from joint differentially private algorithms in the case of large games (Section~\ref{sec:privacy}).

\subsection{Efficiency theorem for cost-minimization games with dynamic population}
We first provide the efficiency theorem for cost-minimization games by extending the proof of Proposition~\ref{prop:smooth_game} to account for the dynamic population. Although our main applications are about mechanisms, we begin with cost-minimization games for pedagogical purposes as the proof is simpler in this case. In Section~\ref{ssec:large-congestion}, we apply this theorem to the application of atomic congestion games (such as routing).
Before presenting our efficiency theorem, we formalize the notion of stability we require from solution sequences of the underlying optimization problem.
\vspace{0.4em}
\begin{definition}[$k$-stable sequence for cost-minimization games]\label{defn:stable} A randomized sequence of solutions $\vec{x}^{1:T}$ and types
$\vec{v}^{1:T}$ is $k$-stable for games if the cumulative (across players) expected number of changes in each individual player's solution or type is at most $k$. More formally, if $k_i(v_i^{1:T},x_i^{1:T})$ is the number of times that player $i$ has $x_i^{t}\neq x_i^{t+1}$ or $v_i^t\neq v_i^{t+1}$, then:
\begin{equation*}
\sum_{i=1}^{n} \E\left[k_i\left(v_i^{1:T},x_i^{1:T}\right)\right]\leq k
\end{equation*}
where the expectation is taken over the randomness in the sequence of types, as well as the associated solution sequence.\footnote{Note that the solution sequence $\vec{x}^{1:T}$ necessarily depends on the sequence of player types $\vec{v}^{1:T}$.}
\end{definition}
\vspace{0.1in}
\begin{theorem}[efficiency theorem for cost-minimization games]\label{thm:poa-cost}
Consider a repeated cost game with dynamic population $\Gamma=(G,T,p)$, such that the stage game $G$ is solution-based $(\lambda,\mu)$-smooth and costs are bounded in $[0,1]$. Suppose that there exists a $k$-stable sequence  $(\vec{v}^{1:T}, \vec{x}^{1:T})$ such that $\vec{x}^t$ is feasible (pointwise) and $\beta$-approximately (in-expectation) optimal for each $t$, i.e. $\E[C(\vec{x}^t; \vec{v}^t)]\leq \beta\cdot \E[\opt(\vec{v}^t)]$. If players use
adaptive learning algorithms then:
\begin{equation*}
\sum_t \E[C(s^t;\vec{v}^t)]\leq \frac{\lambda \beta}{1-\mu}\sum_t \E[\opt(\vec{v}^t)] + \frac{n}{1-\mu}\sqrt{ T\cdot \left(\frac{k}{n}+1\right)\cdot \ln(N T)}
\end{equation*}
\end{theorem}
\input{proof-poa-cost}

\begin{remark}
The above result smoothly degrades if the players use adaptive learning algorithms with worse guarantees. More concretely, if their adaptive regret guarantee for time interval $[\tau_1,\tau_2)$ is: $\E[R(\tau_1,\tau_2)]\leq C_R (\tau_2-\tau_1)^{q}$ for some constant $C_R$ and some 
$q\in[0,1]$, the theorem would still hold with the last term replaced by $\frac{n}{1-\mu} \cdot C_R T^{q} (k+1)^{1-q}$. This can be proved by replacing the use of Cauchy-Schwartz inequality in the proof with the more general H\"older inequality and the same technique can be also applied to all our later results. For ease of presentation, we limit the discussion about use of weaker adaptive learning algorithms to this remark.
\end{remark}

\subsection{Efficiency theorem for mechanisms with dynamic population}
We now describe the efficiency meta-theorem for mechanisms with dynamic population that we use in our main application: simultaneous multi-item auctions (Sections~\ref{ssec:matching}~and~\ref{ssec:large-markets}). This meta-theorem again reduces proving efficiency guarantees of learning outcomes with dynamic population to showing existence of stable and approximately optimal solution sequences. Since, in mechanisms, the relevant players for a solution are those that are allocated a resource in the corresponding allocation, we first define an allocation-based analog of the stability definition \ref{defn:stable}. If a player $i$ is not allocated any resource in an allocation $x$, we denote its allocation as the empty allocation: $x_i=\emptyset$.
\vspace{0.4em}
\begin{definition}[$k$-stable sequence for mechanisms]\label{defn:stable_mechanisms}
A randomized sequence of allocations $\vec{x}^{1:T}$ and types $\vec{v}^{1:T}$ is $k$-stable for mechanisms if the cumulative (across players) expected number of changes in the allocation or type of individual players with nonempty allocation is at most $k$, i.e., if $\kappa_i(v_i^{1:T},x_i^{1:T})$ is the number of times that player $i$ has $x_i^t\neq x_i^{t-1}$ or ($x_i^t\neq \emptyset$ and $v_i^t\neq v_i^{t-1}$) then:
\begin{equation*}
\sum_{i=1}^{n} \E\left[\kappa_i\left(v_i^{1:T},x_i^{1:T}\right)\right]\leq k
\end{equation*}
\end{definition}
Observe that unlike the definition of $k_i(v_i^{1:T},x_i^{1:T})$ in Definition \ref{defn:stable}, $\kappa_i(v_i^{1:T},x_i^{1:T})$ does not account for changes in the type of players that are not currently allocated an item in solution $x_i^t$.

Before providing the efficiency meta-theorem for mechanisms, we stress a property that needs to be satisfied by the mechanism for the efficiency guarantee to hold. This asks that a player who is not allocated any item is getting zero utility, and that players always bid in a way that gives them nonnegative utility. This holds in most auction formats such as first-price, second-price and hybrid auctions under a no-overbidding-assumption. \footnote{A notable exception to the property is any all-pay auction as losing bidders pay their bid and therefore obtain negative utility despite not being allocated anything.}
\vspace{0.1in}
\begin{assumption}\label{ass:indiv} Any player $i$ who is not allocated a resource has zero utility, i.e. $u_i(s_i^*(v_i,\emptyset),s_{-i})=0$; the utility is also always nonnegative, i.e. $u_i(s;v_i)\geq 0$ for any strategy used by the players.
\end{assumption}
This assumption posits that players with no items in the feasible allocation will have no regret against deviating strategy that attempts to ``win'' the empty allocation. We now provide the meta-theorem for mechanisms.

\vspace{0.1in}
\begin{theorem}[efficiency theorem for mechanisms]\label{thm:improved-poa-mech}
Consider a repeated mechanism with dynamic population $\M=(M,T,p)$, such that the stage mechanism $M$ is solution-based $(\lambda,\mu)$-smooth, satisfies Assumption
\ref{ass:indiv}, 
and utilities are in $[0,1]$. Suppose that there exists a $k$-stable sequence $(\vec{v^{1:T},\vec{x}^{1:T})}$ such that $\vec{x}^t$ is feasible (pointwise) and $\beta$-approximately optimal (in-expectation) for each $t$, i.e. $\beta\cdot \E[W(\vec{x}^t; \vec{v}^t)]\geq \E[\opt(\vec{v}^t)]$.
If players use adaptive learning algorithms
 then:
\begin{equation*}
\sum_t \E[W(s^t; \vec{v}^t)]\geq \frac{\lambda}{\beta\max\{1,\mu\}}\sum_t \E[\opt(\vec{v}^t)] - \sqrt{ T\cdot m\cdot (k+m)\cdot \ln(N T)}
\end{equation*}
where $m$ is such that for any feasible allocation $x$, $|\{i: x_i\neq \emptyset\}|\leq m$.
\end{theorem}
\input{proof-improved-poa-mech.tex}

\begin{remark}
\vspace{0.1in}
The bounds presented in Theorems~\ref{thm:poa-cost} and \ref{thm:improved-poa-mech} have a logarithmic dependence on the time horizon $T$, which makes them ineffective when the overall time $T$ is very long. However, this  can be replaced by a logarithmic dependence on $T_{\max}$ at the expense of having $\max(p,1/T_{\max})$ instead of just $p$ in the bounds, assuming players achieve regret that is no worse than algorithms that restart learning
after at most $T_{\max}$
rounds; with $T_{\max}\sim 1/p$ then $\log(T)$ can be replaced by $\log(1/p)$. 
\end{remark}

%% file: proof-poa-cost.tex
% !TEX root=PoA-general.tex
\begin{proof}
In a dynamic population game, the underlying optimization problem is changing over time. Therefore the classical smoothness analysis described in the proof of Proposition \ref{prop:smooth_game} requires a stronger learning property. To deal with this, we define time-dependent deviating strategies that are related to the underlying optimization problem of the round. We then use the adaptive learning property to show that the players do not regret any sequence of such deviating strategies.

Let $s_i^{\star,t}$ be the deviation $s_i^\star(v_i^t,x_i^{t})$ defined by the smoothness property and $s_i^{
\star,1:T}$ be the sequence of these deviations. Let $K_i
$ be the number of rounds that $s_i^{\star,t}\neq s_i^{\star,t+1}$ and $r_i(s_i^{\star,1:T}, s^{1:T}; \vec{v}^{1:T})$ the regret that player $i$ has compared to selecting $s_i^{\star,t}$ at every round, i.e.:
\begin{equation}
r_i(s_i^{\star,1:T}, s^{1:T}; \vec{v}^{1:T})=\sum_{t=1}^{T} \left(c_i(s^t; \vec{v}^t)- c_i(s_i^{\star,t},s_{-i}^t;\vec{v}^t)\right).
\end{equation}
For shorthand, we denote this with $r_i^\star$ in this proof. 
Observe that since $s_i^{\star,t}$ is uniquely determined by $v_i^t$ and $x_i^t$, $K_i$ is a random variable that is equal to the $k_i(v_i^{1:T},x_i^{1:T})$ of Definition \ref{defn:stable}, for each instantiation of the sequences $\vec{v}^{1:T}$ and $\vec{x}^{1:T}$.

Let $\tau_{i,r}$ be the round that the strategy $s_i^{\star,t}$ of player $i$ changes for the $r$-th time.
For any period $[\tau_{i,r},\tau_{i,r+1})$ that the strategy  $s_i^{\star,t}$
is fixed, adaptive learning guarantees that the player's regret for this strategy is bounded by
\begin{equation}
\label{one-period-regret}
R_i(\tau_{i,r},\tau_{i,r+1}) \le  \sqrt{(\tau_{i,r+1}-\tau_{i,r}
)\ln(NT)},
\end{equation}
 Summing over the $K_i+1
$ periods $[\tau_{i,r},\tau_{i,r+1})$ in which the strategy for player $i$ is fixed and using the Cauchy-Schwartz inequality, we can bound the total regret of each $i$:
\begin{equation}
\label{one-player-regret}
r_i^\star
\le 
\sqrt{(K_i
+1) \sum_{r=1}^{K_i
+1}(\tau_{r+1}-\tau_r)\ln(NT)}=
\sqrt{(K_i
+1) T\ln(NT)},
\end{equation}
Thus for each instantiation of $\vec{x}^{1:T}$ and $\vec{v}^{1:T}$, we have:
\begin{equation}
\label{eq:regret}
\sum_{t=1}^{T} c_i(s^t; \vec{v}^t) = \sum_{t=1}^{T} c_i(s_i^{\star,t},s_{-i}^t;\vec{v}^t) + 
r_i^\star \leq \sum_{t=1}^{T} c_i(s_i^{\star,t},s_{-i}^t;\vec{v}^t) +
\sqrt{(K_i+1) T\ln(NT)},
\end{equation}
Adding over all players, and using the smoothness property, we get that
$$
\sum_t C(s^t; \vec{v}^t) \le \lambda\sum_t C(\vec{x}^{t};\vec{v}^t) + \mu \sum_t C(s^t; \vec{v}^t) +\sum_i \sqrt{(K_i+1) T \ln(N T)}.
$$
By Cauchy-Schwartz,
$\sum_i \sqrt{(K_i+1) T \ln(N T)}\leq \sqrt{n \cdot T\cdot \ln(NT)\cdot \sum_{i=1}^{n}(K_i+1)}$.
Taking expectation over the allocation and valuation sequence and using the $\beta$-approximate optimality and Jensen's inequality:
$$
\textstyle{\sum_t \E[C(s^t; \vec{v}^t)] \le \lambda \beta\sum_t \E[\opt(\vec{v}^t)] + \mu \sum_t \E[C(s^t; \vec{v}^t)] +n\cdot %C_R 
\sqrt{T \ln(N T)\left(1+\frac{1}{n}\sum_{i=1}^n\E[K_i
]\right)}.}
$$
By the $k$-stability of the sequence, we have that $\sum_{i=1}^{n}E[K_i ]\leq k$. By re-arranging we obtain the claimed bound.
\end{proof}

%% file: proof-improved-poa-mech.tex
% !TEX root=PoA-general.tex
\begin{proof}
Let $s_i^{\star,1:T}$ be defined exactly as in the proof of Theorem \ref{thm:poa-cost} and $r_i(s_i^{\star,1:T}, s^{1:T}; \vec{v}^{1:T})$, or $r_i^{\star}$ as shorthand, be defined similarly as:
$$
r_i(s_i^{\star,1:T}, s^{1:T}; \vec{v}^{1:T})=\sum_{t=1}^{T} \left(u_i(s_i^{\star,t},s_{-i}^t;\vec{v}^t)- u_i(s^t; \vec{v}^t)\right)
$$
For any period $[\tau_{r},\tau_{r+1})$ that the strategy  $s_i^{\star,t}$ is fixed, adaptive learning guarantees that the player's regret for this strategy is bounded by
\begin{equation*}
R_i(\tau_r,\tau_{r+1}) \le\sqrt{(\tau_{r+1}-\tau_r)\ln(NT)},
\end{equation*}
Moreover, if in period $r$, $x_i^t=\emptyset$, then by Property \ref{ass:indiv} we have that: $R_i(\tau_r,\tau_{r+1})\le 0$. Thus, if we denote with $X_{i,r}$ the indicator of whether in period $r$, $x_i^t=\emptyset$, we obtain:
\begin{equation*}
R_i(\tau_r,\tau_{r+1}) \le\sqrt{X_{i,r}^2(\tau_{r+1}-\tau_r)\ln(NT)},
\end{equation*}
Summing over the $K_i+1$ periods in which the strategy is fixed and using the Cauchy-Schwartz inequality, we can bound the total regret of each $i$:
\begin{align*}
r_i^\star
~=~& \sum_{r=1}^{K_i+1} \sqrt{X_{i,r}} \cdot \sqrt{X_{i,r}(\tau_{r+1}-\tau_r)\ln(NT)}
~\le~\sqrt{\sum_{r=1}^{K_i+1}X_{i,r}}\cdot \sqrt{\sum_{r=1}^{K_i+1}X_{i,r}(\tau_{r+1}-\tau_r)\ln(NT)}
\end{align*}
Let $Y_i^t=1_{\{x_i^t\neq\emptyset\}}$.
Then observe that:
\[\sum_{r=1}^{K_i+1}X_{i,r}(\tau_{r+1}-\tau_r) = \sum_{t=1}^{T} Y_i^t.\]
Replacing in the previous inequality, summing over all players and using Cauchy-Schwartz:
\begin{align*}
\sum_{i=1}^{n} r_i^\star~\le~& \sum_{i=1}^{n}\sqrt{\sum_{r=1}^{K_i+1}X_{i,r}}\cdot \sqrt{\sum_{t=1}^{T}Y_i^t\ln(NT)}
~\le~ \sqrt{\sum_{i=1}^{n}\sum_{r=1}^{K_i+1}X_{i,r}}\cdot \sqrt{\sum_{i=1}^{n}\sum_{t=1}^{T}Y_i^t\ln(NT)}
\end{align*}
Since each $\vec{x}^t$ is a feasible allocation: $\sum_{i=1}^{n} Y_i^t\leq m$. Hence, $\sum_{i=1}^n \sum_{t=1}^T Y_i^t\leq mT$.  Moreover:
\begin{equation*}
\sum_{i=1}^{n} \sum_{r=1}^{K_i+1}X_{i,r} ~\leq~ 
\sum_{i=1}^n \sum_{r=1}^{K_i}X_{i,r}+\sum_{i=1}^n X_{i,K_i+1}
~=~\sum_{i=1}^n \sum_{r=1}^{K_i}X_{i,r}+\sum_{i=1}^n Y_{i}^T
~\leq~ m+\sum_{i=1}^n \sum_{r=1}^{K_i}X_{i,r}
\end{equation*}
Now observe that for each instance of $(v^{1:T}, x^{1:T})$:  $\sum_{r=1}^{K_i}X_{i,r}\leq \kappa_i(v^{1:T},x^{1:T})$, since the latter summation sums all changes in type or allocation ranging from $r=1$ to $K_i$, such that the allocation $x_i^{\tau_r}$ in the period right before the $r$-th change is non-empty. This is at most the set of changes that are accounted in $\kappa_i(v^{1:T},x^{1:T})$. It is an inequality as there could be an index $r$ at which both a type and an allocation is changing and the summation only accounts it once, while $\kappa_i(v^{1:T},x^{1:T})$ counts it twice, or there could be changes where $x_i^{t}\neq x_i^{t-1}$ and $x_i^{t}\neq \emptyset$, which are not accounted in the above, but are accounted in $\kappa_i(v^{1:T},x^{1:T})$.
Combining all the above we get:
\begin{align*}
\sum_{i=1}^{n} r_i^\star
~\le~&\sqrt{m+\sum_{i=1}^n\kappa_i(v^{1:T},x^{1:T})}\cdot \sqrt{m T\ln(NT)}
\end{align*}
By the no-regret property of each player, for each instance of $\vec{x}^{1:T}$ and $\vec{v}^{1:T}$, we have:
\begin{equation*}
\sum_{t=1}^{T} u_i(s^t; \vec{v}^t) ~\geq~ \sum_{t=1}^{T} u_i(s_i^{*,t},s_{-i}^t;\vec{v}^t) - r_i^\star
\end{equation*}
Adding over all players, and using the smoothness property and the bound on the sum of regrets, we get that
$$
\sum_t \sum_{i}u_i(s^t; \vec{v}^t) ~\ge~ \lambda\sum_t W(\vec{x}^{t};\vec{v}^t) - \mu \sum_t \rev(s^t) -\sqrt{m+\sum_{i=1}^{n}\kappa_i(v^{1:T},x^{1:T})}\cdot \sqrt{m T\ln(NT)}
$$
Taking expectation over the allocation and valuation sequence and using the $\alpha$-approximate optimality and Jensen's inequality:
$$
\sum_t \sum_{i}\E[u_i(s^t; \vec{v}^t)] \ge \frac{\lambda}{\beta}\sum_t \E[\opt(\vec{v}^t)] - \mu \sum_t \E[\rev(s^t)] -\sqrt{m+\sum_{i=1}^{n}\E[\kappa_i(v^{1:T},x^{1:T})]}\cdot \sqrt{m T\ln(NT)}.
$$
By re-arranging and using the fact that $W(s^t;\vec{v}^t)=\sum_{i}u_i(s^t;v^t)+\rev(s^t)$ and that $\rev(s^t)\leq W(s^t;\vec{v}^t)$ (since utilities are non-negative by Assumption 
\ref{ass:indiv}), we obtain the claimed bound.
\end{proof} 

%% file: matching.tex
%!TEX root=write-up.tex

In this section we offer direct arguments to show the existence of stable solution sequences and hence good efficiency results for the case of simultaneous first-price auctions with unit-demand bidders. Our method is based on a combination of using the greedy algorithm and rounding the input parameters. 
\label{ssec:matching}
We consider a repeated mechanism with dynamic population $\mathcal{M}=(M,T,p)$, where the stage mechanism 
consists of simultaneous first price auctions with $n$ unit-demand bidders (matching markets) and $m$ items. Recall from Section \ref{ssec:applications} that the value of each bidder for any item is either $0$ or at least a constant $\rho>0$ (for example, a cent of a dollar) and that the bidders are only allowed to bid multiples of $\zeta \rho$ for a small constant $\zeta>0$ (so the bidding space is discrete). To apply our main theorem, Theorem~\ref{thm:improved-poa-mech}, we need two properties: i) that the mechanism is solution-based smooth, and ii) that there exists a relatively stable sequence of approximately optimal solutions for the optimization problem. 

\subsection{Stability via layered greedy algorithm.}
In order to be able to apply our meta-theorem from the previous section, we need to also show that there exists a sequence of approximately optimal solutions that is relatively stable to changes in the population. When bidders are unit-demand then the underlying optimization problem is finding a matching of maximum value between bidders and items. Unfortunately, the optimal algorithm is not stable to changes in the population since a single change in one player's valuation may the allocation of all other players in the optimal solution as the new optimal solution may come from an augmenting path.

To obtain a stable and approximately optimal solution sequence, we use a layered version of the greedy algorithm. The greedy matching algorithm initially does not allocate any item to any player, and subsequently considers item valuations $v_i^t(j)$ in decreasing order and assigns item $j$ to player $i$ %,
if, when $v_i^t(j)$ is considered, neither item $j$ nor player $i$ are matched. To make this algorithm more stable we define the \emph{layered-greedy 
matching algorithm}, which works as follows. Recall that $\rho>0$ denotes the smallest non-zero value that a player has for any item. For a positive $\epsilon \le 1/3$, we round each player's value down to the closest number of the form $\rho(1+\epsilon)^{\ell}$ for some integer $\ell$, and run the greedy algorithm with these rounded values. It is well known that the greedy algorithm guarantees a solution that is within a factor of 2 to optimal. We lose an additional factor of $(1+\epsilon)$ by working with the rounded values.

The greedy algorithm may have many ties and we resolve ties in a way to make the output stable. In particular, among all the player-item pairs with the same rounded value at round $t$, we break ties in favor of pairs that were matched in the previous round $t-1$. Note that again this neither affects anything in the mechanism nor makes any assumption on how the mechanism breaks ties, it is just inside the proof to show the existence of an approximately optimal stable solution sequence. 

We now provide the key ingredient for our efficiency guarantee, a lemma showing that the above algorithm provides a sequence of approximately optimal and stable solutions; and therefore such a sequence exists for the underlying optimization problem. We will use the stability definition for mechanisms (Definition \ref{defn:stable_mechanisms}).
\vspace{0.1in}
\begin{lemma}[Stability via layered-greedy matching]\label{lem:greedy_layered}
For any $\epsilon>0$, there exists a sequence of solutions for the underlying matching problem such that a) the total welfare at any round is near-optimal:
$W(x^t;v^t)\geq \frac{1-\epsilon}{2}\opt(v^t)$ and b) the sequence is $k$-stable for mechanisms (as in Definition \ref{defn:stable_mechanisms}) with
$k=2m\cdot (1+\cdot p\cdot T) \cdot \log_{(1+\epsilon)}(1/\rho)$.
\end{lemma}
\input{proof-greedy-layered}

\subsection{Smoothness with discrete bidding spaces.} 
We now show that this mechanism with the discrete bidding space is smooth for the settings we consider. This section focuses on auctions with unit-demand buyers which is a special case of \emph{submodular} player valuations. A valuation function $v(S)$ for subset of items $S$ is submodular if for all sets $A\subset B$ item $j$ we have
$$
v(A\cup\{j\})-v(A)\ge v(B\cup\{j\})+v(B).
$$
$(1/2,1)$-smoothness of the simultaneous first price auction with submodular bidders and continuous bids was known by Syrgkanis and Tardos \cite{Syrgkanis2013}. A simple modification of the result of \cite{Syrgkanis2013} shows that if the discretization is fine enough, then the mechanism is approximately $(1/2,1)$ solution-based smooth. We provide 
the more general result for submodular valuations, as we re-use this fact in Section~\ref{ssec:large-markets}, where we consider more general valuations. Since the techniques are similar to the ones used in \cite{Syrgkanis2013}, we defer this proof to Appendix \ref{sec:app-smoothness}.

\vspace{0.1in}
\begin{lemma}[Smoothness of simultaneous first price auction with discrete bidding space]\label{lem:xos_smooth}
The simultaneous first price mechanism where players are restricted to bid on at most $d$ items and on each item submit a bid that is a multiple of $\zeta
\cdot \rho$, is a solution-based $\left(\frac{1}{2}-\zeta
,1\right)$-smooth mechanism, when players have submodular valuations, such that all marginals are either $0$ or at least $\rho$ and such that each player wants at most $d$ items, i.e. $v_i(S)=\max_{T\subseteq S:|T|=d}v(T)$. 
\end{lemma}

\subsection{Efficiency guarantee}
We now provide the efficiency guarantee that has many nice properties. First, it is parametric; we get an extra factor of $2$ due to the use of greedy algorithm in the proof but, other than that, the additional error goes to $0$ as $p\rightarrow 0$. Second, the turnover probability can be very high without big loss in efficiency as it does not have any dependence on the number of players or items, depends only the range of item valuations. In particular, even if a constant fraction of the players is changing at every round then we still do not see much loss in efficiency, which makes the price of anarchy guarantee meaningful even with high player turnover.

\vspace{0.1in}
\begin{theorem}[Main theorem for first-price auctions with unit-demand bidders ]\label{thm:matching_markets}
In simultaneous first-price auctions with dynamic population and unit-demand bidders, assume that the average optimal welfare in each round is at least $m\rho$, that is $\frac{1}{T}\sum_{t=1}^T\E[\opt(\vec{v}^t)]\geq m \rho$ (all items can be allocated to someone for the minimum marginal value). If players use adaptive learning algorithms such that Assumption~\ref{ass:indiv} holds and $T\geq \frac{1}{p}$, it holds
:
\begin{equation*}
\textstyle{
\sum_t \E[W(s^t; \vec{v}^t)]\geq \prn*{\frac{1-2\zeta}{4}-\sqrt[3]{p\cdot 16\log(1/\rho)\ln(NT)/\rho^2}}\sum_t \E[\opt(\vec{v}^t)]
}
\end{equation*}
where $N$ is the number of different strategies considered by a player.
\end{theorem}
\input{proof-sim-fpa-matching}

%% file: proof-greedy-layered.tex
\begin{proof}
The solutions in the theorem come by applying the layered-greedy matching algorithm with parameter $\epsilon>0$ (to be set in the end of the section). The approximation result holds as we lose a factor of $2$ due to the greedy algorithm and a factor of $(1+\epsilon)$ due to the layers. Since $\frac{1}{1+\epsilon}>1-\epsilon$, result (a) follows.

To show the stability let $\ell(v_i^t(j))$ be the highest integer $\ell$ such that $v_i^t(j)\ge \rho (1+\epsilon)^{\ell-1}$, i.e., the rounded down version of $v_i^t(j)$ is  $\rho (1+\epsilon)^{\ell(v_i^t(j))-1}$, which we call the layer of this value.
For example,
any value in the range $[\rho,\rho(1+\epsilon))$ is in layer $1$. Let $\ell^t(j)$ denote $\ell(v_i^t(j))$ if  item $j$ is assigned to player $i$ at time $t$, and let $\ell^t(j)=0$ if item $j$ is not assigned at time $t$.
We use the potential function
$$\Phi(x^t)=\sum_j\ell^t(j)$$
to show stability. As all values are upped bounded by $1$, the number of possible values that the potential function can take is $ m \log_{(1+\epsilon)} (1/\rho)$.

The crux of the proof relies in the following steps. We first show that changes in assignments of non-departing players correspond to increases in the potential function. Moreover, the potential function can only decrease due to departures: when a player assigned to item $j$ leaves at time $t$, this immediately decreases the potential function by $\ell^t(j) \le \log_{(1+\epsilon)} (1/\rho)$. Ignoring changes due to one round of increasing the potential, the aggregate increase in the potential function is the same as the aggregate decrease. Hence, the expected number of changes is upper bounded by bounding the expected decrease in the potential function. This argument is formalized below.

The allocation of the solution can change only due to a turnover: either a player that holds an item in the greedy solution departs and leaves her previously assigned item open for current players, or a new player arrives and gets assigned to an item. Every time that a player that holds an item departs, she leaves that item temporarily free in the greedy solution. Unless this is the last time that the item has a holder in the greedy solution, at some point (either at the same round or in the future), some player $i$ will get assigned to this item. Due to the layered version of the greedy algorithm, this increases the potential function by at least $1$. If player $i$ was previously assigned to another item, this item becomes free and some other player may move to that item by increasing again the potential function by at least $1$. As a result, the total increase in the potential function,
caused by someone getting an unassigned item is associated to at most $2$ changes. The case where the item remains unassigned for all the future rounds, contributes in total at most $m$ extra changes in allocation (over the whole time horizon).

Now consider the scenario where an arriving player misplaces another player from the greedy solution (instead of getting an unassigned item). For this to happen, her rounded value is higher than the one of the current owner; hence, the potential function increases by $1$.  This affects the allocation of the previous holder of the item who may either cease being allocated or replace a player in another item again increasing the potential function by at least $1$ (and this may propagate across subsequently misplaced players as well). As a result, each increase in the potential function that is caused by someone getting a previously assigned item contributes at most $2$ changes in the allocation of other players. 

Combined with the previous point, the total number of changes is at most:
\begin{equation}\label{eq:bound_on_changes}
    \sum_i K_i(x^{1:T})\leq 2\cdot \text{Total Increase in $\Phi$}+m
\end{equation}

Since the potential function can  only get integer non-negative values and is bounded by $m\cdot \log_{(1+\epsilon)}(1/
\rho)$, and taking into account end-game effects, the total increase in potential function is:
\begin{equation}\label{eq:bound_on_increase}
\text{Total Increase in $\Phi$} \leq \text{Total Decrease in $\Phi$} + m\cdot \log_{(1+\epsilon)}(1/
\rho).
\end{equation}

We are therefore left to bound the expected total decrease in the potential function. This can only happen when a player among the ones that hold items departs. Each such player departs with probability $p$ and hence the expected number of such players departing at each round is at most
$m\cdot p$ (which is independent of the number of players and depends only on the number of items). Whenever this happens, the potential function can decrease by at most $\log_{(1+\epsilon)}(1/\rho)$ since this is the maximum layer the corresponding item can be in. As a result, the expected decrease in the potential function is at most: \begin{equation}\label{eq:bound_on_decrease}
    \En\brk*{\text{Total Decrease in $\Phi$}}\leq p\cdot m\cdot T\cdot \log_{(1+\epsilon)}(1/\rho)
\end{equation}
The lemma follows by combining $\eqref{eq:bound_on_changes}$, $\eqref{eq:bound_on_increase}$, and $\eqref{eq:bound_on_decrease}$.
\end{proof}

%% file: proof-sim-fpa-matching.tex
% !TEX root=matching.tex

\begin{proof}We apply our efficiency meta-theorem (Theorem \ref{thm:improved-poa-mech}) with $x^{t}$ being the allocation of the layered-greedy algorithm with parameter $\epsilon>0$ to be defined later in the proof (note that we never run this algorithm, it is just inside the proof and therefore it is fine to define the parameter ex post). To do so, we use the $(1/2-\zeta,1)$ solution-based smoothness of the first price auction with discrete bidding space established in Lemma \ref{lem:xos_smooth} and the stability of the solution sequence produced by the layered-greedy algorithm established in Lemma \ref{lem:greedy_layered}. Using that $pT>1$, we therefore obtain: 
\begin{align*}
\sum_t \E[W(s^t; \vec{v}^t)]&\geq \frac{(1-2\zeta)(1-\epsilon)}{4}\sum_t \E[\opt(\vec{v}^t)]\\&\qquad \qquad-\sqrt{T\cdot m\cdot (2m\cdot(1+pT))\log_{(1+\epsilon)}(1/\rho)+m)\cdot \ln(N T)}
\\
&\geq \frac{(1-2\zeta)(1-\epsilon)}{4}\sum_t \E[\opt(\vec{v}^t)] -
2mT\cdot\sqrt{p\cdot \frac{\log(1/\rho)}{\log(1+\epsilon)}\cdot \ln(N T)}
\end{align*}
We now optimize $\epsilon$ to minimize the extra error added due to turnover and layering by equalizing the contributions of regret term and error due to the layering. To facilitate computations, we lower bound the optimum in the first term by $m\rho T$ due to the assumption we made on the optimal solution. Assuming that $\epsilon<1$ (which we will verify later), the denominator of the regret term can also be lower bounded by $\epsilon/2$ due to Taylor expansion. Setting $\epsilon=\sqrt[3]{p\cdot 128\log(1/\rho)\ln(NT)/\rho^2}$, the two terms become equal. Using this, the result holds as the extra loss in the factor is at most $2\cdot\epsilon/4$; note that if $\epsilon>1$ then the result is trivially true as the right hand side is a negative quantity while the left hand side is nonnegative by Property \ref{ass:indiv}.
\end{proof}

\begin{remark}
\vspace{0.1in}
For the regret term involving the turnover probability $p$ to become of the order of $\eta$, we need $p\approx \frac{\rho^2\cdot \eta^3}{\log(NT)\cdot\log(1/\rho)}$ which allows a constant churn of turnover at every round.
\end{remark}

%% file: privacy.tex
% !TEX root=write-up.tex
In this section we use the notion of differential privacy for constructing stable sequences needed by our main Theorems \ref{thm:poa-cost} and \ref{thm:improved-poa-mech}. Differential privacy offers a general framework to find solutions that are close to optimal, yet more stable to changes in the input than the optimum itself. To guarantee privacy, the output of the algorithm is required to depend only minimally on any player's input. This is exactly what we need in our framework.

\subsection{Background on differential privacy.}\label{sub:privacy}

Differential privacy has been developed for databases storing private information for a population. A database $D\in \V^n$ is a vector of inputs,
one for each user. Two databases are \emph{$i$-neighbors} if they differ just in the $i$-th coordinate, i.e., only in the input of
the $i$-th user. If two databases are $i$-neighbors for some $i$, they are called \textit{neighboring databases}.
Dwork et al. \cite{Dwork:2006:CNS:2180286.2180305} define an algorithm as \emph{differentially private} if one user's information has little influence on the outcome.

In the setting of a game or mechanism the outcome for player $i$ clearly should depend on player $i$'s input (her claimed valuation, or source destination pair), so cannot be differentially private.
The notion of \emph{joint differential privacy} has been developed by Kearns et al. \cite{DBLP:conf/innovations/KearnsPRU14} to adapt differential privacy to such settings. We use $\X$ to denote the set of possible outcomes for one player, so an algorithm in this context is a function $\A: \V^n\rightarrow \X^n$. The algorithm is jointly differentially private, if for all players $i$, the output for all other players is differentially private in the input of player $i$. More formally:
\vspace{0.4em}
\begin{definition}[\cite{DBLP:conf/innovations/KearnsPRU14}]\label{definition:joint_differential}
An algorithm $\A: \V^n\rightarrow \X^n$ is $(\epsilon,\delta)$- jointly differentially private if for every $i$, for every pair of $i$-neighbors $D,D'\in \V^n$, and for every subset of outputs $S\subseteq \X^{n-1}$.
\begin{equation}\label{eqn:joint-privacy}
Pr[\A(D)_{-i}\in S]\leq \exp(\epsilon)Pr[\A(D')_{-i}\in S]+\delta
\end{equation}
If $\delta=0$, we say that $\A$ is $\epsilon$-jointly differentially private.
\end{definition}

Over the last years there have been a number of algorithms developed that solve problems close to optimally in a differentially private way; see the book of Dwork and Roth \cite{DworkRothBook} for a survey. In this paper, we take advantage of such algorithms, including the  algorithms for solving matching problems \cite{hsu2016private} and finding socially optimal routing \cite{DBLP:journals/corr/RogersRUW15}.

\subsection{Stability via differential privacy}
We now show that differentially private solutions imply stability. This is the main building block for efficiency guarantees.

\vspace{0.1in}
\begin{lemma}[Stable sequences via privacy]\label{lem:privacy_sequence}
Suppose there exists an algorithm $\A:\V^n\rightarrow \Delta(\X^n)$ that is $(\epsilon,\delta)$-jointly differentially private, takes as input a valuation profile $\vec{v}$ and outputs a distribution of solutions such that a sample from this distribution is feasible with probability $1-\gamma$, and is $\beta$-approximately efficient in expectation  (for $0\leq \epsilon\leq 1/2$, $\beta>1$, $\delta>0$, and $0<\gamma<1$).\footnote{$\gamma$ is used here locally to match the notation used in differential privacy
and not related to the $\gamma$ of the efficiency loss due to regret error in our overall parametric guarantee parametric guarantee.}
Consider the sequence $\vec{v}^{1:T}$ of valuations produced in a repeated cost-minimization game with dynamic population $\Gamma=(G,p,T)$ or mechanism with dynamic population $\mathcal{M}=(M,T,p)$. There exists a randomized sequence of solutions $\vec{x}^{1:T}$ for the sequence $\vec{v}^{1:T}$, such that for each $1\le t \le T$, $\vec{x}^t$ conditional on $\vec{v}^t$ is an $\beta$-approximation to $\opt(\vec{v}^t)$ in expectation and the joint randomized sequence $(\vec{v}^{1:T},\vec{x}^{1:T})$ is $k$-stable (as in either Definition~\ref{defn:stable} or Definition~\ref{defn:stable_mechanisms}) with $k=pnT(1+  n(2\epsilon+2\gamma+\delta))$.
\end{lemma}

To prove this main lemma, we require two auxiliary lemmas. First, we bound the total variation distance between the outputs of a differentially private algorithm on two inputs that differ only on one coordinate (the type of one player). Total variation distance is a general measure for the distance between distributions. For two distributions $\mu$ and $\eta$ on some finite probability space $\Omega$ the following are two equivalent versions of the total variation distance:
\begin{equation}\label{eqn:tv}
d_{tv}(\mu,\eta)=\frac{1}{2}\|\mu-\eta\|_1 = \max_{A\subset \Omega}(\mu(A)-\eta(A)),
\end{equation}
where in the 1-norm in the middle we think of $\mu$ and $\eta$ as a vector of probabilities over the possible outcomes.

\begin{lemma}\label{lem:privacy_variation_dist}
\vspace{0.1in}
Suppose that $\mathcal{A}:\mathbf{V}^n\rightarrow \Delta(\X^n)$ is an $(\epsilon,\delta)$-joint differentially private algorithm with failure probability 
$\gamma$ (for $0\leq \epsilon\leq 1/2$
, $\delta>0$, and $0<\gamma<1$) that takes as input a valuation profile $v$ and outputs a distribution over feasible solutions $\rsolv$. Let $\rsolv$ and $\rsolv'$ be the algorithm's outputs on two inputs $v$ and $v'$ that differ only in coordinate $i$.
Then we can bound the total variation distance between $\rsolv_{-i}$ and $\rsolv_{-i}'$ by $d_{tv}(\rsolv_{-i},\rsolv'_{-i})\le(2\epsilon+\delta)$.
\end{lemma}
\begin{proof}
Condition \eqref{eqn:joint-privacy} of joint differential privacy guarantees that if we let $S\subseteq \X_{-i}^n$ be a subset of possible solutions for players other than $i$, and 
$\rsolv_{-i}(S)$ and $\rsolv'_{-i}(S)$ be the probability that the two distributions assign on $S$, then for any $S$:
$\rsolv_{-i}(S)\leq e^{\epsilon}\cdot\rsolv'_{-i}(S)+\delta$. Since $\epsilon\leq 1/2$, we can use the bound $e^{\epsilon} \le 1+2\epsilon$ to get that $\rsolv_{-i}(S) - \rsolv'_{-i}(S) \leq 2\epsilon \rsolv'_{-i}(S)+\delta\leq 2\epsilon +\delta$. Thus by the second definition of the total variation distance in Equation \eqref{eqn:tv} we get that $d_{tv}(\rsolv_{-i},\rsolv'_{-i})\leq 2\epsilon+\delta$.
%\Halmos
\end{proof}

Second, we use a simple lemma from basic probability theory.

\begin{lemma}[Coupling Lemma]\label{coupling_lemma}
\vspace{0.1in}
Let $\mu$ and $\eta$ be two probability measures over a finite set $\Omega$. There is a coupling $\omega$ of $(\mu,\eta)$, such that if the random variable $(X,Y)$ is distributed according to $\omega$, then the marginal distribution on $X$ is $\mu$, the marginal distribution on $Y$ is $\eta$, and
$$\mathbb{P}\brk*{X\neq Y}=d_{tv}(\mu,\eta), $$
\end{lemma}

\begin{proof}[Proof of Lemma \ref{lem:privacy_sequence}.]
Suppose that $\mathcal{A}:\mathbf{V}^n\rightarrow \Delta(\X^n)$ is an $(\epsilon,\delta)$-joint differentially private algorithm as described in the definition of the lemma. The differentially private algorithm fails with probability $\gamma$. We denote with $\rsolv$ the output distribution over solutions for an input $v$, where we use the optimal solution in the low probability event that the algorithm fails.
(Equivalently $\mathcal{A}$ could be a randomized algorithm and $\rsolv$ its implicit distribution over solutions).

Let $\rsolv^1,\ldots,\rsolv^T$, be the sequence of distributions output by $\mathcal{A}$ when run on a deterministic sequence of valuation profiles $v^1,\dots,v^T$  with the modification described in the paragraph above. To simplify the discussion we assume that only one player changes valuation at each time-step $t$. Essentially we break every transition from time-step $t$ to $t+1$ into many sequential transitions where only one player changes at every time step, and then deleting the solutions from the resulting sequence that correspond to the added steps. Thus the number of steps within this proof should be thought as being equal to $n\cdot p\cdot T$ in expectation.

By Lemma~\ref{lem:privacy_variation_dist}, we know that the total variation distance of two consecutive distributions without the modification of replacing failures with the optimal solution is at most $2\epsilon+\delta$. Since, by the union bound, the probability that any of the two consecutive runs of the algorithm fail is at most $2\gamma$, we can show that the total variation distance of the latter modified output is at most $2\epsilon+\delta+2\gamma$, i.e. for any $t\in [T]$: $d_{tv}(\rsolv_{-i}^{t+1},\rsolv_{-i}^{t})\leq 2\epsilon+\delta+2\gamma$ (see Lemma \ref{lem:modification} for a formal proof).

We can turn the sequence of distributions  $\rsolv^1,\ldots,\rsolv^T$ into a distribution of sequences of allocations $x^{1:T}$ by coupling the randomness used to select the solutions in different distributions $\rsolv^t$. To do this, we take advantage of the coupling lemma from probability theory (Lemma \ref{coupling_lemma}). If at step $t$ no player changes values, then $\rsolv^t=\rsolv^{t+1}$, and we select the same outcome from the two distributions, so we get $\mathbb{P}\brk*{x^t_{-i}\neq x^{t+1}_{-i}}=0$.

Now consider a step in which a player $i$ changes her private type $v_i$. We use  Lemma \ref{coupling_lemma} to couple $x^{t+1}_{-i}$ and $x^{t}_{-i}$ so that\footnote{One can think of it as sampling $x^{t+1}$ conditional on $x^t$ and assuming the joint distribution of $x^t$ and $x^{t+1}$ is as prescribed by the coupling lemma applied to $\sigma^t$ and $\sigma^{t+1}$. This is to address concerns that $x^t$ is already coupled with $x^{t-1}$ in the previous step.}
\begin{equation}\label{prob-bound}
\mathbb{P}[x^{t+1}_{-i}\neq x^{t}_{-i}]=d_{tv}(\rsolv^{t+1}_{-i},\rsolv^t_{-i})\leq 2\epsilon+\delta+2\gamma.
\end{equation}
Note that this couples the $i$th coordinate $x_i^{t+1}$ and $x_i^t$ in an arbitrary manner, which is fine, as we assumed that the valuation of player $i$ changes at this step.

We have defined a probability distribution of sequences $x^{1:T}$ for every fixed sequence of valuations $v^{1:T}$. We extend this definition to random sequences of valuation in the natural way adding the distribution of valuations $v^{1:T}$.

We claim that the resulting random sequences of  (valuation,solution) pairs satisfies the statement of the theorem: the $\beta$-approximation follows by the guarantees of the private algorithm and by the fact that we use the optimal solution when the algorithm fails.
Next we argue about the stability of the sequence. 
Consider a player $i$, and the distribution of her sequence $(v_i^{1:T},x_i^{1:T})$. In each step $t$ her valuation $v_i^t$ changes with probability $p$, contributing $pT$ in expectation to the number of changes.
In a step $t$ when some other value $j\neq i$ changes, we use \eqref{prob-bound} to bound the probability that $x^t_i\neq x^{t+1}_i$ by $2\epsilon+\delta+2\gamma$.
Thus any change in the value of some other player $j$ contributes at most $(2\epsilon+2\gamma+\delta)$ to the expectation of the number of changes for player $i$. The expected number of such changes in other values is $(n-1)pT$ over the sequence, showing that  $$\mathbb{E}\brk*{K_i}=pT+(n-1)pT(2\epsilon+2\gamma+\delta)\leq pT(1+n(2\epsilon+2\gamma+\delta)).$$ 
Summing over players, we obtain the lemma.
\end{proof}

\begin{lemma}\label{lem:modification}
Let $q$ and $q'$ be the output of an $(\epsilon,\delta)$-joint differentially private algorithm with failure probability $\gamma$, on two valuation profiles $v$ and $v'$ that differ only in coordinate $i$. Let $\sigma$ and $\sigma'$ be the modified output where the outcome is replaced with optimal outcome when the algorithm fails. Then:
\begin{equation*}
d_{tv}(\sigma,\sigma')\leq 2\epsilon+\delta+2\gamma
\end{equation*}
\end{lemma}
\begin{proof}
Consider two random coupled random variables $y,y'$ that are implied by Lemma \ref{coupling_lemma} applied to distributions $q$ and $q'$, such that $y\sim q$ and $y'\sim q'$ and $\mathbb{P}\brk*{y\neq y'}=d_{tv}(q,q')\leq 2\epsilon+\delta$ (by Lemma~\ref{lem:privacy_variation_dist}). Now consider two other random variables $x$ and $x'$ where $x=y$ except for the cases where $y$ is an outcome of a failure in which case $x$ is equal to the welfare optimal outcome and similarly for $x'$ and $y'$. Obviously: $x \sim \sigma$ and $x'\sim \sigma'$, thus $(x,x')$ is a valid coupling for distributions $\sigma$ and $\sigma'$. Thus if we show that $Pr[x\neq x'] \leq 2\epsilon+\delta+2\gamma$, then by properties of total variation distance $d_{tv}(\sigma,\sigma')\leq Pr[x\neq x']\leq 2\epsilon+\delta+2\gamma$, which is the property we want to show.

\newcommand{\fail}{\ensuremath{\text{fail}}}
Let $\fail$ be the event that either $y$ or $y'$ is the outcome of a failed run of the algorithm. Then by the union bound $\mathbb{P}\brk*{\fail}\leq 2\gamma$. Thus we have:
\begin{align*}
\mathbb{P}\brk*{x\neq x'} =~& \mathbb{P}\brk*{x\neq x'\mid{\neg \fail}}\cdot\mathbb{P}\brk*{\neg \fail} +\mathbb{P}\brk*{x\neq x' \mid \fail}\cdot\mathbb{P}\brk*{\fail}\\
\leq~&  \mathbb{P}\brk*{x\neq x'\mid \neg \fail}\cdot\mathbb{P}\brk*{\neg \fail} +2\gamma\\
=~&  \mathbb{P}\brk*{y\neq y'\mid\neg \fail}\cdot\mathbb{P}\brk*{\neg \fail} +2\gamma\\
\leq~& \mathbb{P}\brk*{y\neq y'} + 2\gamma \leq d_{tv}(q,q') + 2\gamma\leq 2\epsilon + \delta + 2\gamma
\end{align*}
This completes the proof of the Lemma.
%\Halmos
\end{proof}

\subsection{Efficiency guarantee for dynamic games via differential privacy}
We can now provide the resulting efficiency guarantee, which we instantiate in the next section to simultaneous auctions in large markets as well as routing games.

\begin{theorem}\label{thm:PoA_with-privacy}
\vspace{0.1in}
Consider a repeated cost game with dynamic population $\Gamma=(G,T,p)$, such that the stage game $G$ is solution-based $(\lambda,\mu)$-smooth and $T\geq \frac{1}{p}$. Assume that there exists an $(\epsilon,\delta)$-joint differentially private algorithm $\A:\V^n\rightarrow \X^n$ with approximation factor $\beta$ that satisfies the conditions of Lemma \ref{lem:privacy_sequence}. If all players use adaptive learning algorithms in the repeated game then the overall cost of the solution is at most:
$$
\textstyle{\sum_t \E[C(\vec{s}^t;\vec{v}^t)] \le \frac{\lambda \beta}{1-\mu}\sum_t \opt(\vec{v}^t) + \frac{nT}{1-\mu}\sqrt{p\big(1+2n( \epsilon+\gamma+\delta)\big) \ln(N T)}}
$$
Similarly for a $(\lambda,\mu)$-smooth mechanism $\mathcal{M}=(M,T,p)$ we obtain:
$$
\sum_t \E[W(s^t; \vec{v}^t)]\geq \frac{\lambda}{\beta \max\{1,\mu\}}\sum_t \E[\opt(\vec{v}^t)] - mT\sqrt{\Big(p+p\cdot\frac{n}{m}\cdot (1+  2n(\epsilon+\gamma+\delta)) \Big)\ln(N T)}
$$
\end{theorem}
\begin{proof}
The proof follows directly by combining Lemma~\ref{lem:privacy_sequence} with Theorems~\ref{thm:poa-cost}~and~\ref{thm:improved-poa-mech} respectively.
\end{proof}

%% file: privacy_applications.tex
%!TEX root=write-up.tex
In this section, we combine Theorem~\ref{thm:PoA_with-privacy} with differentially private algorithms to obtain enhanced efficiency guarantees for simultaneous-item auctions as well as guarantees for congestion games. The results of this section require the game to be \emph{large enough} for the privacy results to apply.

\subsection{Improved efficiency for simultaneous-item auctions in large markets}
\label{ssec:large-markets}
\input{large-markets} 

\subsection{Asymptotic optimality under participation uncertainty}\label{ssec:asymptotic_uncertainty} 

We now show that, if there is some amount of participation uncertainty at every round then the efficiency guarantees can be significantly improved for some classes of auctions when the game is sufficiently large, and result in asymptotic optimality when the turnover is also sufficiently small. In order to do so, we use the model of Feldman et al. \cite{FeldmanILRS_STOC16} and extend their efficiency guarantee to dynamic-population games. In this model, at every round, each player may fail to participate with probability $q$ (which can be thought as a small constant). Note that this is different to the turnover probability $p$ that determines whether the player switches their type; each player stays in the game in expectation with unchanged type for $1/p$ rounds and participates in expectation in $(1-q)/p$ rounds. 

We denote by $z^t$ the random variables determining the players who participate: $z_i^t=1$ if player $i$ participates at round $t$ and $0$ otherwise. The welfare at round $t$ with player strategies $s^t$ is correspondingly denoted by $W(s^t,z^t;\vec{v}^t)$ (implicitly assuming that players $i$ who didn't show up may have had a strategy $s_i^t$ also), and the optimum by $\opt(z^t;\vec{v}^t)$. We note that both are random variables; the optimum's randomness comes
from the dynamic population and the randomness of participation uncertainty  $z^t$ in each round, while the welfare of the game outcome $W(s^t,z^t;\vec{v}^t)$   also involves the randomness in the strategies of the players.

The auction format we consider is a large-supply extension of second-price auction, which we refer as \emph{uniform-price} auction. In particular, by the above large market assumptions (Assumption~\ref{ass:large_markets_assumptions})  there are $\widetilde{m}$ distinct items, each with a supple of $B$; therefore there are $m=\widetilde{m}\cdot B$ total items. Each player $i$ places a bid  $b_{ij}$ simultaneously for each distinct item $j\in[\widetilde{m}]$ and appears in the round with probability $1-q$. Among the players that appear, each item $j$ is awarded to the players with the highest $B$ bids and each of them pays a uniform bid equal to the bid of the $(B+1)$-th bidder that appeared.  Uniform-price auctions are natural extensions of first and second price auctions in the large market setting. In particular Theorem \ref{thm:large_markets} and Remark \ref{rem:second-prize} extend to markets with item supply $B$, assuming the price charged is the $B$th and $(B+1)$th prize respectively. Participation uncertainty improves both efficiency bounds by a factor of 2. We illustrate this focusing on the version with $(B+1)$th price.

\begin{theorem}[Asymptotic optimality for large markets with participation uncertainty]\label{thm:large_markets_uncertainty}
\vspace{0.1in}
 In simultaneous uniform-price auctions with dynamic population and $d$-demand bidders with gross-substitute valuations, if Assumption~\ref{ass:large_markets_assumptions} holds, players fail to show up with probability $q$ and use adaptive learning algorithms such that Assumption~\ref{ass:indiv} holds and $T\geq \frac{1}{p}$, as $p\rightarrow 0$ and $B\rightarrow \infty$, the resulting social welfare is asymptotic optimal. Concretely, it holds that, for any $\eta>0$:
$$
 \sum_t\E\Big[W(s^t, z^t;\vec{v}^t)\Big]\geq \Big((1-\zeta)(1-\eta)-\frac{4}{\rho\sqrt{q(1-q)B}}-\frac{1}{\rho}\sqrt{p\cdot (3+2n\epsilon)\ln(NT)}\Big)\sum_t\E\Big[\opt(z^t;\vec{v}^t)\Big]
$$
where $\epsilon=\frac{12\rho^3}{\eta^3 B}\cdot\Big( 360\sqrt{2}\log\big(\frac{90n\rho^2}{\eta^2}\big)^{5/2}\log(4mn^2)+4\Big)$ and $N$ is the number of strategies considered by a player, when the supply $B$ is large enough, given all other parameters, to ensure that $\epsilon<1$.
\end{theorem}
\begin{proof}[Proof sketch]
We now provide the key ideas of the proof; the full proof is deferred to Appendix~\ref{app:asymptotic_uncertain_arrivals}. If there was no uncertainty in which player shows up, then we can directly obtain a smoothness-like inequality that has the $B$-th bid instead of the actual price, which is the $(B+1)$-th bid (Lemma~\ref{lem:large-smooth-utility}); this is similar to $(1,1)$-smoothness and therefore leads to a $(1-\eta)$ multiplicative loss in efficiency. The uncertainty combined with the largeness of the game enables us to show that the difference between the $B$-th and $(B+1)$-th bid is in expectation small (Lemma~\ref{lem:B-B+1}); this leads to the $\frac{4}{\rho\sqrt{q(1-q)B}}$ term in the bound. For each subset of players who may show up, we obtain a stable close-to-optimal solution to the sequence of subproblems involving this subset using differential privacy (similar to Theorem~\ref{thm:large_markets}) that leads to the last term in the bound. Finally, applying Bayesian extension \cite{RoughgardenEC12,Syrgkanis12}, 
we combine the bounds for the subproblems with the expected difference between $B$-th and $(B+1)$-th bids leading to the final guarantee.

For the regret term involving the turnover probability to become order of $\eta$, we again need $p\approx \frac{\rho^5\cdot \eta^5 \cdot B}{n \log(NT)\polylog(m,n,1/\eta,\rho)}$. For the second to-last term to become of the order of $\eta$, we need the supple to be at least $B\approx \frac{1}{\eta^2\rho^2 \cdot q(1-q)}$. The above imply the claim that as $p\rightarrow 0$ and $B\rightarrow \infty$, we have asymptotic optimality.
\end{proof}

\begin{remark}
\label{rem:B-prize}
\vspace{0.1in}
Although we presented the result for uniform-price auctions using the $(B+1)$th prize, the version of uniform-prize used by \cite{FeldmanILRS_STOC16}, the same proof idea also extends to the version of uniform-price auction using the $B$th bid, resulting in a similar bound social welfare, showing asymptotic optimality, assuming $T\geq \frac{1}{p}$, as $p\rightarrow 0$ and $B\rightarrow \infty$. 
\end{remark}

\subsection{Efficiency of congestion games with dynamic population}
\label{ssec:large-congestion}
\input{large-congestion}

%% file: large-markets.tex
% !TEX root=privacy_applications.tex
We first apply the guarantee of the previous section to provide an efficiency result that improves on Theorem~\ref{thm:matching_markets} in two ways: the guarantee is better by a factor of $2$ and the result allows the players to have a broader class of valuations; in particular, they need to satisfy the gross substitute property (Definition~\ref{def:gross_substitutes}). On the negative side, the result of this section 
is effective only when the supply for each item is relatively large while Theorem~\ref{thm:matching_markets} applies even when the supply for each item is just one.

\begin{definition}[Gross-substitute valuations]\label{def:gross_substitutes}
\vspace{0.1in}
For a price $p$ let $p(A)=\sum_{j \in A} p_j$ denote the total price, and let $\omega(p)$ denote the player's most desirable set of goods, that is, let $\omega(p)=\arg\max_A v(A)-p(A)$. The valuation satisfies the gross substitutes condition if for every pair of price vectors $(p, p')$ such that $\forall$ items $j$  $p_j\leq p'_j$ and for every set of goods $S\in \omega(p)$ if $S'\subseteq S$ satisfies $p_j'=p_j$ for every $j\in S'$ then there is a set $S^*\in\omega(p')$ with $S'\subseteq S^*$.
\end{definition}
In order to obtain a stable solution sequence, we apply the differentially private algorithm \emph{PAlloc} of Hsu et al. \cite{hsu2016private} and make the following assumptions required for its efficiency guarantee.
 \begin{assumption}[Large markets assumptions]\label{ass:large_markets_assumptions}
\vspace{0.1in} Recall that $\widetilde{m}$ is the number of distinct items and $B$ is the supply of each item. We posit the following large market assumptions:
\begin{enumerate}
    \item The total number of items $m=\widetilde{m}\cdot B$ is greater than the number of players $n$, i.e. $m\geq n$. 
    \item Each player is interested in at most one copy of each item and is $d$-demand, i.e. gets $0$ additional value from obtaining more than $d$ items.
    \item The average optimal welfare in each round is at least $m\rho$, that is $\frac{1}{T}\sum_{t=1}^T\E[\opt(\vec{v}^t)]\geq m \rho$ (value if all copies can be allocated to someone for the minimum marginal value $\rho$).
\end{enumerate}
\end{assumption}
We now provide the privacy and approximation guarantee (Lemma~\ref{lem:privacy_efficiency_guarantee_palloc}) from \cite{hsu2016private} which we use in the our main efficiency guarantee (Theorem~\ref{thm:large_markets}).
\begin{lemma}[Theorems 20 and 21 in \cite{hsu2016private}]\label{lem:privacy_efficiency_guarantee_palloc}
\vspace{0.1in}
Let $0<\alpha<n/m$ and $\vec{x}$ be the allocation computed by $PAlloc(\alpha/3,\alpha/3,\epsilon)$. Then, if Assumption~\ref{ass:large_markets_assumptions} holds and on top of that the supply for each distinct item is at least:
$B\geq \frac{12E'+3}{\alpha}$ where $E'=\frac{360\sqrt{2}}{\alpha^2\epsilon}\Big(\log(\frac{90n}{\alpha^2}\Big)^{5/2}\log(\frac{4m}{\gamma})+1$ then the algorithm is $(\epsilon,0)$ differentially private and, with probability $1-\gamma$, the allocation satsfies
$$W(\vec{x})\geq \opt(\vec{v})-\alpha m.
$$
\end{lemma}

\begin{theorem}[Main theorem for large markets with gross-substitute $d$-demand bidders]\label{thm:large_markets} In simultaneous first-price auctions with dynamic population and $d$-demand bidders with gross-substitute valuations (Definition~\ref{def:gross_substitutes}), if Assumption~\ref{ass:large_markets_assumptions} holds, players use adaptive learning algorithms such that Assumption~\ref{ass:indiv} holds and $T\geq \frac{1}{p}$, it holds that, for any $\eta>0$:
$$
\sum_t \E[W(s^t; \vec{v}^t)]\geq \frac{1-2\zeta}{2}\cdot (1-\eta)\Big(1-\frac{1}{\rho}\sqrt{p\cdot (3+2n\epsilon)\ln(NT)}\Big)\sum_t \E[\opt(\vec{v}^t)].
$$
where $\epsilon=\frac{12\rho^3}{\eta^3 B}\cdot\Big( 360\sqrt{2}\log\big(\frac{90n\rho^2}{\eta^2}\big)^{5/2}\log(4mn^2)+4\Big)$ and $N$ is the number of strategies considered by a player, when the supply $B$ is large enough, given all other parameters, to ensure that $\epsilon<1$.
\end{theorem}
\begin{proof}
By Lemma~\ref{lem:xos_smooth}, the mechanism is $(\frac{1}{2}-\zeta,1)$-smooth.  Moreover, by Lemma~\ref{lem:privacy_efficiency_guarantee_palloc} and by condition 3 of Assumption~\ref{ass:large_markets_assumptions}, with probability $1-\gamma$, the allocation $\vec{x}$ of $PAlloc(\alpha/3,\alpha/3,\gamma)$ has welfare satisfying:
$\beta\cdot W(\vec{x})\geq \opt(\vec{v})
$ for $\beta=\frac{1}{1-\alpha}$. We set $\gamma=\frac{1}{2n}$. Then, since PAlloc is $(\epsilon,0)$-joint differentially private, when the supply requirement in Lemma~\ref{lem:privacy_efficiency_guarantee_palloc} is met, Theorem~\ref{thm:PoA_with-privacy} implies:
$$
\sum_t \E[W(s^t; \vec{v}^t)]\geq \frac{1-2\zeta}{2}\cdot (1-\frac{\alpha}{\rho})\sum_t \E[\opt(\vec{v}^t)] - mT \sqrt{p+p\cdot \frac{n}{m}\cdot (2+2n\epsilon)\ln(NT)}.
$$
Setting $\epsilon=\frac{12}{\alpha^3 B}\cdot\Big( 360\sqrt{2}\log\big(\frac{90n}{\alpha^2}\big)^{5/2}\log(4mn^2)+4\Big)$ satisfies the supply requirement in Lemma~\ref{lem:privacy_efficiency_guarantee_palloc}. Setting $\eta=\frac{\alpha}{\rho}$ and using conditions 1 and 3 of Assumption~\ref{ass:large_markets_assumptions} concludes the final bound.
\end{proof}

\begin{remark} 
\vspace{0.1in}
For the regret term involving the turnover probability $p$ to become of the order of $1-\eta$, we need $p\approx \frac{\rho^5\cdot \eta^5 \cdot B}{n \log(NT)\polylog(m,n,1/\eta,\rho)}$ which still allows a near-constant churn of turnover at every round when $B=\Theta(n)$.
\end{remark}
\begin{remark} \label{rem:second-prize}
\vspace{0.1in}
Although we presented the result for first-price auctions, the same result extends to second-price auctions under no-overbidding assumption. The second-price auction is not a smooth mechanism, it is only weakly $(1,0,1)$-smooth in the sense of \cite{Syrgkanis2013}, or is also $(1,1)$-smooth in the sense of \cite{RoughgardenJACM} assuming that players do not bid above their value as shown by Christodoulou et al. \cite{ChrKovSch16}. These alternate smoothness definitions also imply a factor of $1/2$ loss of social welfare at Nash. This can also be extended to solution-based smoothness, and then used in the framework above to obtain the same bounds as Theorem \ref{thm:large_markets} above. In the next part, we show how arrival uncertainty helps enhance the bounds for the natural extension of second price of item auction with many copies of each item. 
\end{remark}

%% file: large-congestion.tex
% !TEX root=write-up.tex
Our final application of privacy-induced stability is atomic congestion games, defined in Section~\ref{ssec:applications}. Rogers et al. \cite{DBLP:journals/corr/RogersRUW15} provide a a jointly differentially private algorithm for finding an optimal solution in congestion games, called  \textit{Private gradient descent algorithm}. They focus on routing games due to the paper's focus on tolls as mediators, but their algorithm works in full generality for any atomic congestion game. The results in this paper rely on the following assumptions for the latency in edge.

\begin{assumption}[Latency assumptions]\label{ass:routing_assumptions}
\vspace{0.1in}
 The latency functions $\ell_e(x)$ are non-decreasing, convex and twice differentiable. The latency on each edge is bounded by $1$, that is, $\ell_e(x)\le 1$. The functions are $L$-Lipschitz, that is $|\ell_e(x)-\ell_e(x')| \le L|x-x'|$ for some parameter $0< L< 1$.
\end{assumption}
Under these assumptions on the setting, Rogers et al.~\cite{DBLP:journals/corr/RogersRUW15} show that their algorithm is approximately optimal and differentially private.

\begin{lemma}[Theorems 4.2 and 4.10 in \cite{DBLP:journals/corr/RogersRUW15}]\label{lem:privacy_large_congestion}\vspace{0.1in}
When Assumption \ref{ass:routing_assumptions} holds, the algorithm of \cite{DBLP:journals/corr/RogersRUW15} is $(\epsilon,\delta)$-joint differentially private and, with probability at least $1-\gamma$, outputs an integer solution $\vec{x}$ that satisfies $$
\E[C(\vec{x};\vec{v})] \le \opt(\vec{v}) +\frac{\sqrt{n}m^{5/4}}{\sqrt{\epsilon}}\cdot\polylog(1/\epsilon, 1/\delta, 1/\gamma, n, m).
$$
\end{lemma}

We start by illustrating 
our technique with linear latencies $\ell_e(x)=a_e x+b_e$. The linear-latency case admits $(5/3,1/3)$ solution-based smoothness:
\begin{lemma}[due to \cite{Christodoulou:2005:PAF:1060590.1060600,DBLP:journals/siamcomp/AwerbuchAE13,Roughgarden:2009:IRP:1536414.1536485}]\label{lem:linear_smooth}
\vspace{0.1in}
Congestion games with linear latencies $\ell_e(x)=a_e x+b_e$ for $a_e,b_e \ge 0$ are $(5/3,1/3)$ solution-based smooth with respect to any solution $\vec{x}$.
\end{lemma} 

Combining the above smoothness result (Lemma~\ref{lem:linear_smooth}) and privacy (Lemma~\ref{lem:privacy_large_congestion}) results with the general efficiency guarantee (Theorem~\ref{thm:PoA_with-privacy}) leads to the following guarantee.

\begin{theorem}[Main theorem for large congestion games with linear latencies]\label{thm:routing}\vspace{0.1in} In linear-latency congestion games with dynamic population, if players use adaptive learning algorithms and $T\geq \frac{1}{p}$, it holds that, for any $\eta>0$:
$$
\sum_t \E[C(\vec{s}^t;\vec{v}^t)] \le \frac{5}{2}\cdot \Big(1+\eta+ \frac{m^2\max_e(a_e+b_e)}{\min_e a_e}\cdot \sqrt{p(1+6n\epsilon)\ln(NT)}\Big)\sum_t \opt(\vec{v}^t)
$$
where $\epsilon=\frac{1}{\eta^2}\cdot\frac{m^{9/2} \cdot \polylog(n,m)}{n}\cdot \Big(\frac{\max_e (a_e+b_e)}{\min_e a_e)}\Big)^2$ and  and $N$ is the number of strategies considered by a player, when $n$ is large enough, given all other parameters,  to ensure that $\epsilon<1$.
\end{theorem}
\begin{proof}
We assume $a_e>0$ for all $e\in E$ and that $b_e\geq 0$. We also scale down all latencies by $nm\max_e(a_e+b_e)$ which makes the loss incurred by each player at most $1$ (needed for the online learning algorithms); this also helps satisfy Assumption~\ref{ass:routing_assumptions}, as it makes the latencies $\ell_e(x)\leq 1$ and $L$-Lipschitz for $L=1/n<1$. As a result, all the additive bounds should be renormalized by $nm\max_e(a_e+b_e)$.
From Theorem~\ref{thm:PoA_with-privacy}, it therefore holds that:
$$
\textstyle{\sum_t \E[C(\vec{s}^t;\vec{v}^t)] \le \frac{\lambda \beta}{1-\mu}\sum_t \opt(\vec{v}^t) + \frac{nT}{1-\mu}\sqrt{p\big(1+2n( \epsilon+\gamma+\delta)\big) \ln(N T)}}\cdot nm\max_e(a_e+b_e)
$$
where $\beta$ is the approximation factor of the joint-differentially private solution. Since in the optimal solution we need to serve all players (and each uses at least one edge), the optimal solution is at least $\opt(\vec{v}^t)\geq \frac{n^2}{m}\min_e a_e$. As a result, to transform the bound in Lemma~\ref{lem:privacy_large_congestion} to a multiplicative bound of $\beta=1+\eta$, it should hold that:
$$
\eta \opt(\vec{v}^t)\geq\eta \frac{n^2}{m}\min_e a_e \geq \frac{\sqrt{n}m^{5/4}}{\sqrt{\epsilon}}%3m\sqrt{n\ln(m/\gamma)} 
\cdot\polylog(1/\epsilon, 1/\delta, 1/\gamma, n, m)\cdot mn\max_e (a_e+b_e).
$$
This is guaranteed by setting $\epsilon=\delta=\gamma =\frac{1}{\eta^2}\cdot\frac{m^{9/2} \cdot \polylog(n,m)}{n}\cdot \Big(\frac{\max_e (a_e+b_e)}{\min_e a_e)}\Big)^2$. The proof concludes by bounding the regret term via a similar argument by $$\frac{\sum_t \opt(\vec{v}^t)}{1-\mu}\sqrt{p(1+2n(\epsilon+\delta+\gamma))\ln(NT)}\cdot \frac{m^2\max_e(a_e+b_e)}{\min_e a_e}.$$
\end{proof}

\begin{remark}
For the regret term involving the turnover probability $p$ to become of order $\eta$, we need 
$$p\approx \frac{\eta^4}{m^{10}\cdot \log(NT)\polylog(m,n)}\cdot \Big(\frac{\min_e a_e}{\max_e(a_e+b_e)}\Big)^3.$$ Note that this only depends on the number of congestible elements (edges) and is independent (up to logarithms) to the number of players, allowing for a near-constant churn of turnover.
\end{remark}

\begin{remark}
\vspace{0.1in}
In the above presentation we used linear latencies but we note that more general latencies would only affect the smoothness factor; as a result, similar techniques can extend the results for polynomial latencies due to \cite{Roughgarden03,Olver,Aland2011} to dynamic population. 
\end{remark}

%% file: app_prelims.tex
%!TEX root = main.tex
\begin{proof}[Proof of Proposition~\ref{prop:smooth_game}]
Let $s^t$ be the strategy vector used by the players at round $t$. Since each player $i$ has low regret subject to the strategy $s^\star_i(x_i,v_i)$,
it holds that $\sum_t c_i(s^t;v_i)\leq \sum_t c_i(s_i^{\star}(v_i,x_i),s_{-i}^t;v_i)+R_i(1,T)$. Adding up these inequalities and using the smoothness property, we obtain:
\begin{align*}
\frac{1}{T}\sum_t C(s^t;\vec{v})&=\frac{1}{T}\sum_t\sum_{i\in [n]}c_i(s^t;v_i) \le
\frac{1}{T}\sum_t\sum_i c_i(s_i^{\star}(v_i,x_i),s_{-i}^t;v_i) +\frac{1}{T}\sum_iR_i(1,T) \\&\leq \lambda C(\vec{x};\vec{v}) + \mu \frac{1}{T}\sum_t C(s^t;\vec{v})+\frac{1}{T}\sum_iR_i(1,T)
\end{align*}
The claimed bound follows by rearranging the terms and since $\frac{1}{T}\sum_iR_i(1,T)\leq \nicefrac{n}{\sqrt{T}}$.
\end{proof}

\begin{proof}[Proof of Proposition~\ref{prop:smooth_mechanism}]
Let $s^t$ be the strategy vector used by players at round $t$. Since each player $i$ has low regret subject to strategy $s^\star_i(v_i,x_i)$, it holds that $\sum_t u_i(s^t;v_i)\geq \sum_t u_i(s_i^{\star}(v_i,x_i),s_{-i}^t;v_i)-R_i(1,T)$. Adding up these inequalities and using the smoothness property, we obtain:

\begin{align*}
\frac{1}{T}\sum_t\sum_{i\in [n]}u_i(s^t;v_i)  &\ge
\frac{1}{T}\sum_t\sum_i u_i(s_i^{\star}(v_i,x_i),s_{-i}^t;v_i)  -\frac{1}{T}\sum_iR_i(1,T) \\
&\geq W(\vec{x};\vec{v}) - \mu \frac{1}{T}\sum_t \rev(s^t)-\frac{1}{T}\sum_iR_i(1,T)
\end{align*}
The claimed bound follows by rearranging the terms and since $\frac{1}{T}\sum_iR_i(1,T)\leq \nicefrac{n}{\sqrt{T}}$.
\end{proof}

%% file: app-smoothness.tex
%!TEX root = main.tex
In this section, we show the smoothness of first-price auction with discrete bid spaces (Lemma~\ref{lem:xos_smooth}); the proof is similar to the one of Syrgkanis and Tardos \cite{Syrgkanis2013} but needs to be adapted to work with discrete bids and is provided therefore for completeness.

\begin{proof}[Proof of Lemma \ref{lem:xos_smooth}]
Consider a valuation profile $v=(v_1,\ldots,v_n)$ for the $n$ players and a bid profile $b=(b_1,\ldots,b_n)$. Each valuation $v_i$ is submodular and thereby also falls into the class of XOS valuations \cite{Lehmann:2001:CAD:501158.501161}, i.e. it can be expressed as a maximum over additive valuations. More formally, for some index set ${\cal L}_i$:
\begin{equation*}
v_i(S) = \max_{\ell\in {\cal L}_i} \sum_{j\in S} a_{ij}^\ell
\end{equation*}
Furthermore, when the player has value for at most $d$ types of items, it can also be shown that for any $\ell\in {\cal L}_i$ at most $d$ of the $(a_{ij}^{\ell})_{j\in [m]}$ will be non-zero. 

Consider a feasible allocation $x=(x_1,\ldots,x_n)$ of the items to the bidders, where $x_i$ is the set of types of items allocated to player $i$ (the latter is feasible if each item is never allocated more than its supply). We assume without loss of generality that each allocation $x_i$ is minimal in the sense that deleting any item from $x_i$ decreases the value for player $i$. Let $m_i$ denote the number of items allocated to player $i$. Observe that we must have $v_i(x_i)\ge \rho m_i$, as the function is submodular, and by minimality, adding items one-by-one, each item must increase the value.

Consider the following deviation $b_i^*(v_i,x_i)$ that is related to the valuation $v_i$ of player $i$ and to allocation $x_i$: Let $\ell^*(x_i) = \arg\max_{\ell\in {\cal L}_i} \sum_{j\in x_i} a_{ij}^\ell$. Then on each item $j\in x_i$ with $a_{ij}^{\ell^*(x_i)}>0$, submit $\left\lfloor\frac{a_{ij}^{\ell^*(x_i)}}{2}\right\rfloor_{\zeta\cdot \rho}$, where we denote with $\left\lfloor \xi\right\rfloor_{\zeta\cdot \rho}$ the highest multiple of $\zeta\cdot \rho$ that is less than or equal to $\xi$. On each $j\notin x_i$, submit a zero bid; this submits at most $d$ non-zero bids. This deviation implies the solution-based smooth property. Let $p_j(b)$ be the lowest winning bid on item $j$, under bid profile $b$. For each $j$, if $p_j(b)<\left\lfloor\frac{a_{ij}^{\ell^*(x_i)}}{2}\right\rfloor_{\zeta\cdot \rho}$, $i$ wins item $j$ and pays $\left\lfloor\frac{a_{ij}^{\ell^*(x_i)}}{2}\right\rfloor_{\zeta\cdot \rho}$. Thus we obtain:
\begin{align*}
u_i(b_i^*(v_i,x_i),b_{-i};v_i)\geq~& \sum_{j\in x_i} \left(a_{ij}^{\ell^*(x_i)}-\left\lfloor\frac{a_{ij}^{\ell^*(x_i)}}{2}\right\rfloor_{\zeta\cdot \rho}\right)\cdot 1\left\{p_j(b)<\left\lfloor\frac{a_{ij}^{\ell^*(x_i)}}{2}\right\rfloor_{\zeta\cdot \rho}\right\}\\
\geq~& \sum_{j\in x_i} \left\lfloor\frac{a_{ij}^{\ell^*(x_i)}}{2}\right\rfloor_{\zeta\cdot \rho}\cdot 1\left\{p_j(b)<\left\lfloor\frac{a_{ij}^{\ell^*(x_i)}}{2}\right\rfloor_{\zeta\cdot \rho}\right\}\\
\geq~& \sum_{j\in x_i} \left(\left\lfloor\frac{a_{ij}^{\ell^*(x_i)}}{2}\right\rfloor_{\zeta\cdot \rho}-p_j(b)\right)\\
\geq~& \sum_{j\in x_i} \left(\frac{a_{ij}^{\ell^*(x_i)}}{2}-\zeta\cdot \rho-p_j(b)\right)\\
=~& \left(\frac{1}{2}\sum_{j\in x_i} a_{ij}^{\ell^*(x_i)}-\zeta\cdot\rho\cdot m_i\right)-\sum_{j\in x_i}p_j(b)\\
\geq~& \left(\frac{1}{2}-\zeta\right)v_i(x_i)-\sum_{j\in x_i}p_j(b)
\end{align*}
Summing over all players and observing that $\rev(b)\geq \sum_{j\in x_i}p_j(b)$, we get the theorem.
\end{proof}

%% file: large-market-smooth.tex
%!TEX root=write-up.tex
In this section, we provide the proof of the asymptotically optimal efficiency guarantee at the presence of participation uncertainty (Theorem~\ref{thm:large_markets_uncertainty}).

For a  participation vector $z$, such that $z_i=1$ if player $i$ shows up and $0$ otherwise, let $x(z)$ denote a feasible solution to the problem with participation $z$ and $SW(x(z),\vec{v})$ be the social welfare of this solution with participant values $\vec{v}$.  We first provide an inequality very similar to $(1,1)$-solution-based smoothness that holds for any such subset of players that end up showing up. In what follows, we use $b_j(k,z)$ to denote the $k$-th highest bid on item $j$ among the players with $z_i=1$, and $u_i(b,z_{-i}, v_i)$ the utility of player $i$ with value $v_i$, bids $b$, where $z_{-i}$ indicates the other participants.

\begin{lemma}\label{lem:large-smooth-utility}
\vspace{0.1in}
For any valuations $\vec{v}$ and any participation vector $\bar z$ there is a bid-vector $b_i^\star(x(\bar z))$ for all bidders $i$ depending on the solution $x(\bar z)$ such that for any bid vector $b$ and any participation vector $z$ the following holds:

$$\sum_i \bar z_i u_i(b_i^\star(x(\bar z))), b_{-i},z_{-i}, v_i) \ge SW(x(\bar z), \vec{v})-\sum_j n_j(x(\bar z))b_j(B, z)$$
where $n_j(x(\bar z))$ is the number of copies of item $j$ allocated by solution $x(\bar z)$.

If the proposed $b_i^\star$ needs to be restricted to only use multiples of $\zeta
\cdot \rho$ then we have 
$$\sum_i \bar z_i  u_i(b_i^\star(x(\bar z)), b_{-i}, z_{-i}, v_i) \ge (1-\zeta)SW(x(\bar z), \vec{v})-\sum_j n_j(x(\bar z)b_j(B, z)$$
\end{lemma}

\begin{proof}
Recall from the proof of Lemma \ref{lem:xos_smooth} (Appendix~\ref{sec:app-smoothness}) that, since valuations are gross-substitute, they can be expressed as a maximum over additive valuations. More formally, for each player $i$ and some index set ${\cal L}_i$:
$$
v_i(S)=\max_{\ell\in {\cal L}_i} \sum_{j\in S} a_{ij}^{\ell}
$$
Consider a feasible allocation  $x(\bar z)=(x_1, \ldots, x_n)$ of items to bidders, with each $x_i$ is minimal, as in the proof of  Lemma \ref{lem:xos_smooth}, where we temporarily dropped the $\bar z$ in the notation for $x_i$ in this proof. Let $\ell$ be the index so $v_i(x_i)=\sum_{j \in x_i} a_{ij}^{\ell}$, and we define $b_{ij}^{\star}(x(\bar z))=a_{ij}^{\ell}$ for all items $j\in x_i$ and 0 otherwise. For a bidder $i$, assume that with bid $b^\star_i(x(\bar z))$ the bidder wins a set of items $x_i'$ among the items in $x_i$. Notice that for all participants with $\bar z_i=1$ we have:
\begin{eqnarray*}
u_i(b_i^\star(x(\bar z)), b_{-i}, z_{-i}, v_i) &\ge & \sum_{j \in x'_i}(a_{ij}^{\ell}-b_j(B, z)) \\
&\ge &\sum_{i \in x_i}(a_{ij}^{\ell}-b_j(B, z)) =v_i(x_i)-\sum_{j \in x_i} b_j(B, z)
\end{eqnarray*}
where the first inequality follows as $v_i(x_i')\ge \sum_{j \in x'_i} a_{ij}^{\ell}$, winning extra times beyond $x_i$ at 0 prize can only improve the value, and
with the extra bid $b_i^\star(x(\bar z))$, the price can be as high as $b_j(B,\bar z)$.
The second inequality follows as for any item $j\in x_i\setminus x'_i$ we must have that the bid $b_{ij}^\star(x(\bar z))\le b_j(B, z)$. 

Now add the utility bounds over all players to get the claimed bound:
\begin{eqnarray*}
\sum_i \bar z_i u_i(b_i^\star(x(\bar z)), b_{-i}, z_{-i}, v_i) &\ge & \sum_i \bar z_i (v_i(x_i)-\sum_{j \in x_i} b_j(B, z)) = SW(x(\bar z),\vec{v})-\sum_j n_j(x(\bar z)) b_j(B, z)
\end{eqnarray*}

If the bidders are restricted to only use multiples of $\zeta
\cdot \rho$ then we use $b_{ij}^\star(x(\bar z))=\zeta
\cdot \rho \lfloor\frac{ a_{ij}^{\ell}}{\zeta\cdot \rho} \rfloor$, and get for a player $i$ participating
$$u_i(b_i^\star(x(\bar z)), b_{-i},z_{-i}, v_i) \ge  v_i(x_i)-\sum_{j \in x_i}(b_j(B,z)-\zeta\rho)$$
summing over all players gives us
\begin{eqnarray*}
\sum_i \bar z_i u_i(b_i^\star(x(\bar z)), b_{-i},  z_{-i}, v_i) & \ge & SW(x(\bar z), \vec{v})-\sum_j n_j(x(\bar z))(b_j(B, z)-\zeta\rho)\\
& \ge & (1-\zeta) SW(x(\bar z), \vec{v})-\sum_j n_j(x)b_j(B, z)
\end{eqnarray*}
where the last inequality follows by the assumption that each item allocated to a player has marginal value at least $\rho$.
\end{proof}

Note that $(1-\zeta,1)$-smoothness of the mechanism would be implied if we had $b_j(B+1, z)$ in place of $b_j(B, z)$ in the above expression as $b_j(B+1, z)$ for $\bar z=z$ is the uniform price charged for each of the $B$ copies of the item in the auction with bids $b$ and participation $z$. 
The main insight in using the above smoothness-like claim is that, due to the uncertainty, the bids $b_j(B,z)$ and $b_j(B+1,z)$ have almost the same distribution as was showed by Lemma 16 in \cite{FeldmanILRS_STOC16}.

\begin{lemma}[Lemma 16 in \cite{FeldmanILRS_STOC16}]\label{lem:B-B+1}\vspace{0.1in}
The expected difference between the $B$-th and $(B+1)$-st bids is at most $$\E_z(b_j(B,z))-\E_z(b_j(B+1,z))\le \frac{4}{\sqrt{q(1-q)B}}.
$$
\end{lemma}
\begin{proof}[Proof of Theorem~\ref{thm:large_markets_uncertainty}]
Let $x^t(z)$ be the allocation returned by PAlloc and $\opt^t(z)$ be the optimum 
for the auction at time $t$ with participants $z$. 

At a high level the proof combines Lemma \ref{lem:large-smooth-utility} used for the solutions $x^t(z)$ and Lemma \ref{lem:B-B+1} with the Bayesian extension of \cite{RoughgardenEC12,Syrgkanis12} that was also used by \cite{FeldmanILRS_STOC16}.
Fix the set of players that appear at any particular round $t$ and call it $z^t$.
For any alternate participation vector $\bar z$, since player $i$ uses adaptive learning algorithms and the allocation of PAlloc is $(\epsilon,0)$-differentially private, applying Lemma~\ref{lem:privacy_sequence} (as in Theorem~\ref{thm:PoA_with-privacy}), it holds:
\begin{align}\label{eq:no_regret_large}
\E(\sum_t z_i^t  u_i(b^t, z^t_{-i}, v_i^t))\geq \E(\sum_t z_i^t & u_i(b_i^\star(x^t(1, \bar z_{-i})), b_{-i}^t, z^t_{-i}), v_i^t) -Reg_i.
\end{align}
where $Reg_i$ is the expected 
regret of player $i$'s learning algorithm, and the expectation is over the randomization by the learning algorithm. 

Next we take expectation over $z^t$ as well as over $\bar z$ on both sides of the inequality. One term on the right hand side will be 
\begin{align}
\E_{z^t,\bar z}( z_i^t  u_i(b_i^\star(x^t(1,\bar z_{-i})), b_{-i}^t, z^t_{-i}, v_i^t))=\E_{z^t,\bar z}( \bar z_i  u_i(b_i^\star(x^t(\bar z)), b_{-i}^t, z^t_{-i}, v_i^t))\label{eq:trick}
\end{align}
where the equation follows noting that the expression on the left doesn't depend on $\bar z_i$, so we can rename $z^t_i$ as $\bar z_i$ as the two have the same distribution, and replacing $1$ by $\bar z_i$ has no effect due to the multiplier $\bar z_i$.

By Lemma~\ref{lem:large-smooth-utility}, for any fixed participation vector $\bar z$,  for all deviating bids $b_i^{\star}(x^{t}(\bar z))$ with valuations~$\vec{v}^t$:
\begin{align}\label{eq:smoothness_cumulative_large}\sum_i \bar z_i u_i(b_i^\star(x^t(\bar z)), b_{-i}^t, z^t_{-i}, v_i^t) & \ge  (1-\zeta)SW(x^t(\bar z), \vec{v^t})-\sum_j n_j^t(x^t(\bar z))b_j^t(B, z^t)\nonumber\\ 
& \ge (1-\zeta)SW(x^t(\bar z),\vec{v^t})-\sum_j B \cdot b_j^t(B, z^t)
\end{align}
Next we sum equation \eqref{eq:no_regret_large} for all players, take expectation over $z^t$ and $\bar z$. Finally, using Eqs~\eqref{eq:trick} and \eqref{eq:smoothness_cumulative_large} for each $\bar z$, and combining with Lemma~\ref{lem:privacy_efficiency_guarantee_palloc}, we obtain:
\begin{align}
\sum_t\sum_i \E[z_i^t  u_i(b^t, z^t_{-i}, v_i^t)]+&\sum_t\sum_j \E\big[B\cdot b_j^t(B, z^t)\big] \nonumber\\ 
& \geq \sum_t (1-\zeta)\E(\opt(\bar z;\vec{v}^t))-(1-\zeta)\alpha mT-\sum_i Reg_i
\end{align}
where the expectation on the left hand side is over the bidding strategies of the players as well as the randomness due to dynamic population and the participation $z^t$, while expectation on the right hand side is over $\bar z$ and the randomness due to dynamic population. 

Combining Theorem~\ref{thm:improved-poa-mech} and Lemma~\ref{lem:privacy_sequence} (as in Theorem~\ref{thm:PoA_with-privacy}), the total expected regret is at most
\begin{align}
    \sum_i Reg_i \le mT\sqrt{\Big(p+p\cdot\frac{n}{m}\cdot (1+  2n(\epsilon+\gamma)) \Big)\ln(N T)}
\end{align}
Now letting $E'=\frac{360\sqrt{2}}{\alpha^2\epsilon}\Big(\log(\frac{90n}{\alpha^2}\Big)^{5/2}\log(\frac{4m}{\gamma})+1$ as in Lemma~\ref{lem:privacy_sequence}, when $B\geq \frac{12E'+3}{\alpha}$, it holds:
\begin{align}
    \sum_t\sum_i \E(z_i^t  u_i(b^t, z^t_{-i}, v_i^t))+\sum_t\sum_j \E\big[B\cdot b_j^t(B, z^t)\big]&\geq \sum_t (1-\zeta)\E(\opt(\bar z;\vec{v}^t))-(1-\zeta)\alpha mT\nonumber\\ &- mT\sqrt{\Big(p+p\cdot\frac{n}{m}\cdot (1+  2n(\epsilon+\gamma)) \Big)\ln(N T)}\label{eq:efficiency_large_last_bid}
\end{align}
Using Lemma~\ref{lem:B-B+1} for the participation vectors $z^t$, the LHS of Eq.~\eqref{eq:efficiency_large_last_bid} can be upper bounded by the expected welfare $\sum_{t} \E(W(s^t,z^t; \vec{v}^t))$ with a small error term, where the expectation is over the dynamic population, participation, as well as randomness in learning 
algorithms as follows:
\begin{align}\label{eq:LHS_effieciency_large_last_bid}
    \sum_t\E\Big[W(s^t,z^t; \vec v^t)\Big]+\frac{4mT}{\sqrt{q(1-q)B}}\geq \E\Big[\sum_t\sum_i z^t_i  u_i(b^t, z^t_{-i}, v_i^t)+\sum_t\sum_j B\cdot b_j^t(B,z^t)\Big].
\end{align}

Similar to the proof of Theorem~\ref{thm:large_markets}, Assumption~\ref{ass:large_markets_assumptions} combined with Eqs~\eqref{eq:efficiency_large_last_bid} and \eqref{eq:LHS_effieciency_large_last_bid} imply:
$$
 \sum_t\E\Big[W(s^t,z^t;\vec{v}^t)
 \Big]\geq \Big((1-\zeta)(1-\frac{\alpha}{\rho})-\frac{4}{\rho\sqrt{q(1-q)B}}-\frac{1}{\rho}\sqrt{p+p\cdot \frac{n}{m}\cdot (2+2n\epsilon)\ln(NT)}\Big)\E\Big[\opt(z^t;\vec{v}^t)\Big]
$$
with $\epsilon=\frac{12}{\alpha^3 B}\cdot\Big( 360\sqrt{2}\log\big(\frac{90n}{\alpha^2}\big)^{5/2}\log(4mn^2)+4\Big)$, concluding the proof.
\end{proof}